\documentclass[aps,pra,reprint,superscriptaddress,nofootinbib]{revtex4-2}
\usepackage{soul}
\usepackage{color}
\usepackage{graphicx}
\usepackage{enumitem}
\usepackage{amsfonts}
\usepackage{amssymb}
\usepackage{amsmath,latexsym,amsthm}
\usepackage{physics}
\usepackage{bm}
\usepackage{xcolor}
\usepackage{diagbox}
\usepackage[breaklinks=true,colorlinks=true,linkcolor=teal,urlcolor=teal,citecolor=teal]{hyperref}
\usepackage[bottom]{footmisc}
\newcommand{\kb}{\mathbf{k}}
\newcommand{\kmax}{k_{\text{max}}}

\newcommand{\wmax}{\Omega_{\text{max}}}

\newcommand{\ad}[0]{\hat{a}^{\dagger}_{\kb}}
\newcommand{\aop}[0]{\hat{a}_{\kb}}
\newcommand{\bop}[0]{\hat{b}}
\newcommand{\bd}[0]{\hat{b}^{\dagger}}
\newcommand{\bxop}[0]{\hat{b}_{x}}
\newcommand{\bxd}[0]{\hat{b}^{\dagger}_{x}}
\newcommand{\byop}[0]{\hat{b}_{y}}
\newcommand{\byd}[0]{\hat{b}^{\dagger}_{y}}
\newcommand{\bdp}[1]{b^{\dagger #1}}
\newcommand{\aopt}[0]{\widetilde{a}_{\kb}}
\newcommand{\adt}[0]{\widetilde{a}^{\dagger}_{\kb}}
\newcommand{\wk}[0]{\Omega_{\mathbf{k}}}
\newcommand{\sigmat}[0]{\widetilde{\sigma}}
\newcommand{\bopt}[0]{\widetilde{b}}
\newcommand{\bdt}[0]{\widetilde{b}^{\dagger}}
\newcommand{\bdtp}[1]{\widetilde{b}^{\dagger #1}}
\newcommand{\tone}[0]{t_{1}}
\newcommand{\ttwo}[0]{t_{2}}
\newcommand{\rmd}{\mathrm{d}}
\newcommand{\eref}[1]{(\ref{#1})}

\newcommand{\ar}{\alpha_{R}}
\newcommand{\ai}{\alpha_{I}}
\newcommand{\zr}{\xi_{R}}
\newcommand{\zi}{\xi_{I}}
\newcommand{\aor}{\alpha_{0R}}
\newcommand{\aoi}{\alpha_{0I}}
\newcommand{\aver}[1]{\left\langle #1 \right\rangle}
\def\Equal{\texttt{=}}

\begin{document}
\title{Detecting quantum vacuum fluctuations of the electromagnetic field}

\author{Aaron R. Malcolm}
\email{aaron.malcolm@warwick.ac.uk}
\author{B. Sharmila}
\email{sharmila.balamurugan@warwick.ac.uk}
\author{Zhi-Wei Wang}
\email{zhiweiwang.phy@gmail.com}
\author{Animesh Datta}
\email{animesh.datta@warwick.ac.uk}
\affiliation{Department of Physics, University of Warwick, Coventry CV4 7AL, United Kingdom.}

\date{\today}

\begin{abstract}
Quantum vacuum fluctuations of the electromagnetic field result in two signatures on a harmonically trapped charged particle: a shift from the natural trap frequency and generation of quantum coherences.
We assess the role of the long-wavelength and rotating-wave approximations in estimating this frequency shift.
We estimate the magnitude of the frequency shift using parameters from a single-electron cyclotron experiment and also demonstrate how the dependence of the frequency shift on the magnetic field of the cyclotron is tied to the rotating-wave approximation. We expect the frequency shift to be observable in future experiments.
We also suggest a possible route to detecting vacuum-generated quantum coherences. 
These experiments should settle the debate on the choice of approximations and gauge in capturing the effect of the quantum vacuum fluctuations.

\end{abstract}

\maketitle

A uniquely quantum aspect of a field are its vacuum fluctuations.
The direct detection of the vacuum fluctuations of the electromagnetic (EM) field via the measurement of the Lamb shift~\cite{LambShift,Bethe,Welton} spurred the development of quantum electrodynamics~\cite{schweber2020}. 
Its macroscopic analogue - the Casimir effect~\cite{CasimirPolder,Casimir1948,Casimir} remains an active area of research~\cite{Stange2021}. 

In this letter, we theoretically identify and quantitatively estimate the extent of the two signatures of vacuum fluctuations of the EM field on a harmonically trapped particle.
The mathematical model is that of a harmonically trapped particle interacting with the vacuum modes of the EM field acting as a bath~Fig.~(\ref{fig:modelsketch}, Inset).
The first signature is a shift in the natural oscillator frequency due to the interaction of the particle with both the resonant and non-resonant EM vacuum modes. 
The second signature is the generation of quantum coherences between the eigenstates of the particle's motion via its interaction with the non-resonant modes of the vacuum EM field. Our results are obtained using a master equation approach which includes both the resonant and non-resonant modes of the EM vacuum field.
This takes us beyond the rotating-wave approximation (RWA), enabling us to identify contributions to the frequency shift that are gauge-dependent against gauge-independent ones.
The role of the RWA has been studied in the context of other vacuum fluctuation phenomena such as vacuum-induced Berry phase~\cite{Liu2011}.

We seek to settle the debate on the choice of proper gauge~\cite{Stokes} and 
fundamental Lagrangians~\cite{gundhi2024} in capturing the effect of a quantum EM vacuum environment. 
The salience of this debate has risen in light of efforts to understand the interface of quantum mechanics and gravity, where experiments remain challenging.
In this paper, we theoretically investigate the detection of quantum EM vacuum signatures in future experiments. 
Our studies of a harmonically trapped charged particle are motivated by single-electron cyclotrons (SECs)~\cite{Gabrielse,Xing2022,Xing2023}.
Relative shifts of $\sim 10^{-10}$ have been detected in the SEC~\cite{Xing2023}.

In order to quantitatively estimate the shifts to the cyclotron frequency in Table~\ref{tab:estimates}, we use the system parameter values considered in~\cite{Xing2023}. However, we note that in order to truly model the SEC \cite{Xing2023}, one must account for additional details including the presence of a cavity that allows only discrete modes of the field vacuum and the interaction of the electron spin with the magnetic field.  

\begin{table}[h]
\begin{tabular}{|c | c | c | c |} 
 \hline
 \diagbox{Approx.}{$\wmax$~~} & $\rule{0cm}{0.5cm}\Omega_{1}$ \Equal $\frac{2 \pi c}{d_{a}}$ & $\Omega_{2}$ \Equal $\frac{c  m  d_{c}  \omega_{c}}{\hbar} $ & $\Omega_{3}$ \Equal $\sqrt{\frac{2 m c^{2} \omega_{c}}{\hbar}}$ \\[5pt]
  \hline
 With RWA & $\rule{0cm}{0.5cm} - 1.1 \times 10^{-11}$ & $-2.0 \times 10^{-11}$ & $-2.0 \times 10^{-11}$ \\[5pt]
 \hline
 Without RWA & $\rule{0cm}{0.5cm}9.4 \times 10^{-15}$ & $9.6 \times 10^{-17}$ & $9.2 \times 10^{-17}$ \\[5pt]
 \hline
\end{tabular}
\caption{Estimates for the relative frequency shift $\Delta \omega_{c}/\omega_{c}$ for different choices of cut-off $\wmax$ with and without the rotating-wave approximation (RWA). The cut-offs $\Omega_{i}$ and the shifts are calculated using the mass of an electron $m$ and the experimental parameters $\omega_{c}$, $d_{a}$ and $d_{c}$ outlined in Fig.~\ref{fig:modelsketch}. The three cut-offs correspond to three choices of the minimum wavelength of the vacuum modes interacting with the electron: (1) the largest amplitude of the electron in the SEC, (2) the de Broglie wavelength of the electron and (3) the zero-point motion of the electron.} 
\label{tab:estimates}
\end{table}
 
In Table~\ref{tab:estimates}, the shift to the cyclotron frequency is estimated without the RWA to be at most $\sim 10^{-15}$. This could become detectable in future experiments with improved sensitivity.
To avoid an ultraviolet divergence in our computation of the shift, we choose an upper cut-off frequency in the EM modes. 
Three different choices consistent with the long-wavelength approximation (LWA) are presented in Table~\ref{tab:estimates}. We note here that the choice of the cut-off introduces significant changes in the frequency shift only when RWA is not considered. In the paper, we demonstrate how these gauge-dependent results depend significantly on the specific high-frequency cut-off chosen for the EM vacuum modes.

The part of the frequency shift that is due to the non-resonant EM modes in our model is similar to the vacuum Bloch-Siegert shift. 
This shift was recently observed in a solid-state cavity QED system~\cite{Li}. 
Detecting it in the SEC~\cite{Xing2023} could provide interesting insights.
The current efforts in improving the sensitivity of this system to search for new physics beyond the Standard Model~\cite{Guellati-Khelifa}
should benefit from our predictions. 

\textit{Model}: We study the interaction of a particle of mass $m$ and charge $q$ in a one-dimensional harmonic potential of frequency $\omega_{c}$, with the vacuum modes of the EM field.
The minimal coupling classical Hamiltonian, $H$ in the Coulomb gauge is
\begin{align}
    \label{eq:Ham1}
    \nonumber H=&\frac{|p_{x} \mathbf{e}_{x} - q A ( \mathbf{r} , t )|^{2}}{2m} + \frac{m \omega_{c}^{2} x^{2}}{2}\\
    &+ \frac{\epsilon_{0}}{2} \iiint_{V} \rmd^{3}\mathbf{r}' \left( |\mathbf{E}(\mathbf{r}',t)|^{2} + c^{2}  |\mathbf{B}(\mathbf{r}',t)|^{2}\right).
\end{align}
where $\mathbf{r}\equiv(x,y,z)$ denotes the position and $\mathbf{p}\equiv(p_{x},p_{y},p_{z})$ the corresponding canonical momentum of the particle. We note that the motion along the $y$- and $z$-axes are constrained, setting $p_{y}=p_{z}=0$. Here $
\mathbf{e}_{x}$ is the unit vector along the $x$-axis, $\mathbf{A}(\mathbf{r}, t)$ is the magnetic vector potential, 
and $\mathbf{E}(\mathbf{r},t)$ (respectively, $\mathbf{B}(\mathbf{r},t)$) is the electric (respectively, magnetic) field. The choice of the Coulomb gauge has implications on the final master equation which will be discussed later.

We neglect the spin of the particle which could have been captured by a $\mathbf{\mu} \cdot \mathbf{B}$ term in Eq.~\eref{eq:Ham1}, where $\mathbf{\mu}$ is the magnetic moment of the charged particle.
A comparison of the magnitude of this term to that of $(q/m) \mathbf{p} \cdot \mathbf{A}$ shows the former to be much smaller so long as $\wk \ll 10^{24} s^{-1}$ (see Appendix \ref{app:spinint} in the Supplementary Material for further details). This is indeed the case in the long-wavelength limit we invoke below.

We quantize Hamiltonian \eref{eq:Ham1} using
\begin{subequations}
\begin{align}
    \label{eq:Afield}
    &\mathbf{\hat{A}}(\mathbf{r},t) = \sum_{\mathbf{k}, \mu} \sqrt{\frac{\hbar}{2 \Omega_{\mathbf{k}} \epsilon_{0} V}} \mathbf{s}_{\mathbf{k}, \mu} \left(\hat{a}_{\mathbf{k}}(t) e^{i \mathbf{k} \cdot \mathbf{r}} + \hat{a}^{\dagger}_{\mathbf{k}}(t) e^{-i \mathbf{k} \cdot \mathbf{r}}  \right),\\
    \label{eq:b}
    &\hat{x}= \sqrt{\frac{\hbar}{2 m \omega_{c}}} \left( \hat{b} + \hat{b}^{\dagger} \right), \quad \text{and} \quad
    \hat{p}_{x} = \frac{\sqrt{\hbar m \omega_{c}}}{\sqrt{2} \, i}\left( \hat{b} - \hat{b}^{\dagger} \right),
\end{align}
\end{subequations}
where $\bop$ and $\bd$ are the ladder operators corresponding to the particle subsystem, satisfying $[\bop,\bd]=1$. The photon annihilation and creation operators for each mode of the field, namely, ($\aop$, $\ad$), are characterised by the wave vector $\kb$ and polarisation vector $\mathbf{s}_{\mathbf{k}, \mu}$, with  $[\aop,a^{\dagger}_{\mathbf{k'}}]=\delta_{\kb,\mathbf{k'}}$. Further, $\Omega_{\kb} = c k $ is the mode frequency with $k=|\kb|$, $\epsilon_{0}$ is the vacuum permittivity and $V$ is an appropriately chosen normalisation volume.

We make another approximation:
\begin{enumerate}[label=(\roman*)]
\setcounter{enumi}{0}
    \item \textit{Weak field limit}, $|\mathbf{A}(\mathbf{r}, t)|^{2} \ll \left\vert \dfrac{2}{q} \: \mathbf{p} \cdot \mathbf{A} (\mathbf{r}, t)\right\vert$,
    \item \label{Ass:LongWave} \textit{Long-wavelength limit}, $e^{\pm i \kb \cdot \mathbf{r}}\approx 1$.
\end{enumerate}
The validity of \ref{Ass:LongWave} is assessed in Appendix \ref{app:lwa} in the Supplementary Material.
Using Assumptions \ref{Ass:WeakField} and \ref{Ass:LongWave}, the Hamiltonian \eref{eq:Ham1} can be rewritten as
\begin{equation}
   \label{eq:Ham}
    \hat{H} = \hat{H}_{\textsc{p}} + \hat{H}_{\textsc{f}} + \hat{H}_{\textsc{i}},
\end{equation} 
where
\begin{subequations}
\begin{align}
\label{eq:Hp}
    &\hat{H}_{\textsc{p}} = \hbar \omega_{c} \left( \bd \bop + \frac{1}{2} \right),\\
    \label{eq:HF}
    &\hat{H}_{\textsc{f}} = \sum_{\kb,\mu} \hbar \Omega_{\kb} \left( \ad \aop + \frac{1}{2} \right) \; \text{and,} \\
    \label{eq:HI}
    &\hat{H}_{\textsc{i}} =  \sum_{\kb,\mu} g_{\kb,\mu} \left(\bop - \bd \right)\left( \aop + \ad \right).
\end{align}
\label{eq:Hfi}
\end{subequations}
Here the subscripts $\textsc{p}$, $\textsc{f}$, and $\textsc{i}$ denote the particle, field and interaction subsystems respectively. The coupling strength
\begin{align}
    \label{eq:gk}
    g_{\kb,\mu} =  \frac{q \hbar}{2 i } \, \sqrt{\frac{\omega_{c}}{\epsilon_{0} m \Omega_{\mathbf{k}} V}} \, (\mathbf{s}_{\mathbf{k}, \mu} \cdot \mathbf{e}_{x}).
\end{align}

As the field consists of a continuum of modes, determining the dynamics given by the full Hamiltonian \eref{eq:Ham} is not tractable in closed form. 
Further, as we are interested only in the dynamics of the particle within the trap, we consider the particle to be an open quantum system interacting with the field.
 The quantum state of the particle $\hat{\sigma}(t)$ at an instant of time $t$ is given by a master equation, obtained by tracing out the field modes of the joint state $\hat{\rho}(t)$ of the particle and the field subsystems at each instant, $\hat{\sigma}(t) = \text{Tr}_{\textsc{f}}[\hat{\rho}(t)]$. 

The master equation is obtained using the Born-Markov approximations listed below:
\begin{enumerate}[label=(\roman*)]
\setcounter{enumi}{2}
    \item \label{Ass:tscale} There exists a separation of timescales between particle and field subsystems;
    \item \label{Ass:weakInt} Weak interaction between the particle and field is assumed. The master equation is therefore found to be second-order in the coupling $g_{\kb,\mu}$;
    \item \label{Ass:eqbm} The field is assumed to be at thermodynamic equilibrium. Therefore, the density matrix of the field can be considered a statistical mixture of the eigenstates of the field Hamiltonian.
\end{enumerate}

\noindent
Being interested in the signatures of quantum vacuum fluctuations, we derive the master equation for the evolution of the quantum state $\hat{\sigma}$ of the particle under the Hamiltonian $\hat{H}$ in Eq.~\eqref{eq:Ham} using Assumptions \ref{Ass:tscale}--\ref{Ass:eqbm}, while considering the field to be in the vacuum state $\langle \ad \aop \rangle = 0$. The salient steps involved are listed in {Appendix \ref{app:MEderiv} in the Supplementary Material.

The resulting master equation is
\begin{align}
    \label{eq:mastereq2}
    \frac{d \hat{\sigma}}{dt} = - &  i \left( \omega_{c} + \Delta_{-} - \Delta_{+} \right) \left[ \bd \bop , \hat{\sigma} \right] \nonumber \\
    + & \Gamma \left( \bop \hat{\sigma} \bd - \frac{1}{2} \bd \bop \hat{\sigma} - \frac{1}{2} \hat{\sigma} \bd \bop \right) \nonumber \\ 
    + & i \Delta_{+} \left( \bd \hat{\sigma} \bd - \bdp{2} \hat{\sigma} - \bop \hat{\sigma} \bop + \hat{\sigma} \bop^{2} \right) \nonumber \\
    + & i \Delta_{-} \left( \bd \hat{\sigma} \bd - \hat{\sigma} \bdp{2} - \bop \hat{\sigma} \bop + \bop^{2} \hat{\sigma} \right) \nonumber \\
    + & \frac{1}{2} \Gamma \left( \hat{\sigma} \bdp{2} - \bd \hat{\sigma} \bd - \bop \hat{\sigma} \bop + \bop^{2} \hat{\sigma} \right), 
\end{align}
where 
\begin{align}
\label{eq:gamma}
\Gamma &= \frac{1}{4 \pi \epsilon_{0}} \, \frac{4 \, q^{2} \omega_{c}^{2} }{3 \, m c^{3}},
\end{align}
and
\begin{align}
\label{eq:deltapm}
\Delta_{\pm} = \frac{\Gamma}{2 \pi \omega_{c}} \left(\pm \wmax - \omega_{c} \ln \left|\frac{\omega_{c} \pm \wmax}{\omega_{c}} \right| \right).
\end{align}
Here $\wmax$ is the cut-off frequency of the EM modes, introduced to avoid the ultraviolet divergences that appear in the computation of relevant particle observables and their dynamics in the presence of vacuum fluctuations of the EM field~\cite{Bethe,Welton}.
It is also known that introducing a cut-off need not lead to a relativistically invariant result~\cite{French}. 
The Compton frequency $\Omega_{\text{Comp}} = 2 \pi c / \lambda_{\text{Comp}} = m c^{2}/\hbar$, commonly sets the upper cut-off in mode frequencies interacting with the vacuum \cite{Bethe,Welton}. 

Our use of the long-wavelength approximation (LWA) \ref{Ass:LongWave} results in  $\wmax=\text{min}(\Omega_{\textsc{lwa}},\Omega_{\text{Comp}})$, where $\Omega_{\textsc{lwa}}$ is the maximum mode frequency assumed when considering $e^{i \kb \cdot \mathbf{r}}\approx 1$. 
An order-of-magnitude estimate gives
\begin{equation}
    \frac{\Omega_{\textsc{lwa}}}{c} < \left(\sqrt{\frac{\hbar}{2m\omega_{c}}}\right)^{-1}  
    \implies \Omega_{\textsc{lwa}} < \sqrt{\frac{2 m c^{2} \omega_{c}}{\hbar}}.
\end{equation}
Here the ground state motion of the charged particle is used to set a limit on the minimum wavelength and hence maximum frequency. In Appendix \ref{app:lwa} in the Supplementary Material, we show that this order-of-magnitude estimate conforms with the results from a detailed time-independent perturbation. 
For typical parameter values from experimental setups of interest (for instance, Fig. \ref{fig:modelsketch}), $\sqrt{2 m c^{2} \omega_{c}/\hbar} \ll \Omega_{\text{Comp}}$.
This allows us to consider different choices of the cut-off frequency such that $\wmax\leqslant\sqrt{2 m c^{2} \omega_{c}/\hbar}$. 

In Eq.~\eref{eq:deltapm}, $\Delta_{\pm}$ comprises two terms, i.e., 
(a) a term linearly dependent on $\Omega_{\text{max}}$, and 
(b) a term with $\ln |\omega_{c}\pm\Omega_{\text{max}}|$. As argued by Bethe \cite{Bethe}, the former vanishes on renormalisation, as it can be reduced to a change in an experimentally irrelevant effective mass. This term is also identified as the computed energy shift due to the presence of vacuum fluctuations on a completely free, untrapped particle. We compute the energy shift of a free, untrapped particle in Appendix \ref{app:FreePart} in the Supplementary Material and find it to be non-zero. However, from symmetry arguments we expect the free particle to be unaffected by vacuum fluctuations. We must therefore subtract away this apparent energy shift in the context of the free particle through renormalisation in order to achieve the correct physics. This unobservable energy shift must then also be subtracted from the un-renormalised energy shift computed for the harmonic oscillator (see Appendix \ref{app:FreePart} in the Supplementary Material for more details). This leads to the shifts in Eq.~\eref{eq:mastereq2} being replaced by their renormalised values
\begin{align}
    \label{eq:deltare}
	\Delta_{\pm}^{\textsc{(r)}} \equiv -\frac{\Gamma}{2 \pi}  \ln \left|\frac{\omega_{c} \pm \wmax}{\omega_{c}} \right| \stackrel{}{\approx}
	\frac{\Gamma}{2 \pi} \left( \ln \left|\frac{\omega_{c}}{ \wmax} \right| \mp \frac{\omega_{c}}{\wmax}\right),
\end{align} 
where the approximation holds for $\wmax \gg \omega_{c}.$

The master equation Eq.~\eqref{eq:mastereq2} is a Redfield equation~\cite{Redfield} which does not guarantee a positive time evolution of the state $\hat{\sigma}$. However, in Appendix \ref{app:validity} in the Supplementary Material, we show that $\hat{\sigma}(t)$ remains positive for time instants up to $t=T_{\text{max}}$, defined as (recalling Eq. \eref{eq:tmax}),
\begin{align}
    \nonumber T_{\text{max}} = \left.\left[\Gamma + \sqrt{\Gamma^{2} + 4 \left(\Delta_{-}^{\textsc{(r)}}\right)^{2}}\right]\middle/ \left[4 \left(\Delta_{-}^{\textsc{(r)}}\right)^{2}\right]\right. \approx \frac{1}{\Gamma}.
\end{align} 
Finally, note that the master equation \eref{eq:mastereq2} has some gauge-dependent terms~\cite{Stokes}. This is discussed in detail later in this letter by comparing with gauge-invariant results obtained under additional assumptions.

The master equation Eq.~\eqref{eq:mastereq2} describes three physical processes: decay, frequency shift and the generation of coherences.  
We attribute the latter two to the presence of the quantum vacuum and discuss them now. 

\textit{Frequency shift}:
The first line in Eq. \eref{eq:mastereq2} shows a shift in the natural particle frequency $\omega_{c}$. After renormalisation, we find the frequency shift to be
\begin{align}
	\label{eq:deltanew}
	\Delta \omega_{c} \equiv \Delta_{-}^{\textsc{(r)}}  - \Delta_{+}^{\textsc{(r)}} =  -\frac{\Gamma}{2 \pi}  \,  \ln \left| \frac{\omega_{c} - \wmax}{\omega_{c} + \wmax}  \right|.
\end{align}
In the limit $\wmax \to \infty$, $\Delta \omega_{c} \to 0$. 
This is in agreement with previous works \cite{Gabrielse,Maclay} which state that the change in the energy levels of the cyclotron (viz. particle in an oscillator) due to the vacuum fluctuations in this limit can be seen as an overall change to the reference energy level. However, we show that when the cut-off takes a finite value, there is a finite shift to the particle frequency. This shift arises due to the presence of vacuum fluctuations as $\Delta \omega_{c}$ is directly proportional to $\langle[\aop,\ad]\rangle,$ as shown in Appendix \ref{app:MEderiv} in the Supplementary Material. 
This can be identified as a `spontaneous' radiative shift in the particle frequency~\cite{Cohen} and an observable signature of the vacuum fluctuations.

Even when we consider the field is not in vacuum, i.e., $\langle \ad \aop \rangle\neq 0$, this shift, $\Delta \omega_{c}$, remains unchanged and is independent of the state of the field. 
In this case, perturbative calculations associate an additional and equal `stimulated' radiative shift to each energy level.
 However, this leaves the particle frequency unchanged as the energy level separation is unaffected by these equal energy shifts. In the open quantum systems approach, this is seen as the contributions to the shift from the stimulated emission and stimulated absorption cancelling each other. Therefore, stimulated processes do not contribute to the shift in trap frequency~\cite{Cohen}.

Expanding Eq.~\eref{eq:deltanew} for $\wmax \gg \omega_{c}$, 
\begin{align}
    \label{eq:deltalimit}
    \Delta \omega_{c} = \frac{ \Gamma \omega_{c}}{\pi \Omega_{\text{max}}} = \frac{1}{4 \pi \epsilon_{0}} \, \frac{4 \, q^{2} \omega_{c}^{3} }{3 \pi \, m c^{3}} \frac{1}{\Omega_{\text{max}},
}.
\end{align}

As this shift depends on the ratio $q^{2} / m$, the ratio is maximum when the chosen particle is an electron. Setting $q=e$ for the electron charge, the relative shift is
\begin{align}
    \label{eq:deltarel}
   \frac{\Delta \omega_{c}}{\omega_{c}} &= \frac{4 \alpha_{\textsc{fs}}}{3 \pi} \, \frac{\hbar \omega_{c}}{m c^{2} } \, \frac{\omega_{c}}{\Omega_{\text{max}}},
\end{align}
where $\alpha_{\textsc{fs}} = e^{2} / (4 \pi \epsilon_{0} \, \hbar c)$ is the fine structure constant. 
The corresponding frequency shift is also obtained using time-independent second-order perturbation in Appendix \ref{app:lwa} in the Supplementary Material.

As the relative shift $\Delta \omega / \omega_{c}$ in Eq.~\eref{eq:deltarel} is directly proportional to $\omega_{c}^{2}$ and inversely to $m$, a high-frequency electron trap,
such as a single-electron cyclotron~\cite{Geonium} is the best experiment for detecting it.

\textit{Coherences:} Lines 3 - 5 of the master equation Eq.~\eref{eq:mastereq2} describe the processes through which quantum coherence is generated 
by the vacuum. This can be seen by considering the time evolution of an off-diagonal element, such as $\sigma_{02}$, of the density matrix $\hat{\sigma}$ of the particle, obtained using Eq. \eref{eq:mastereq2}.
\begin{align}
	\label{eq:sigma02}
	\dot{\sigma}_{02} &= \left(2i\left(\omega_{c} + \Delta_{-} - \Delta_{+} \right) - \Gamma \right) \sigma_{02} \nonumber \\
	&+ \left(\sqrt{3} \Gamma - i 2 \sqrt{3} \Delta_{-} \right) \sigma_{04} + \sqrt{3} \Gamma \sigma_{13} \nonumber \\
	&-  \left(\sqrt{2} i \Delta_{-} + \sqrt{2} i \Delta_{+} + \frac{1}{\sqrt{2}}\Gamma \right) \sigma_{11} \nonumber \\
	&+ \sqrt{2} \left(i \Delta_{-} - \frac{1}{2}\Gamma \right) \sigma_{22} + i \sqrt{2} \Delta_{+} \sigma_{00}. 	
\end{align}
Here $\sigma_{m n} \equiv \bra{m} \hat{\sigma} \ket{n},$ $m,n$ denoting the eigenstates of the harmonic oscillator.
 From Eq.~\eref{eq:sigma02}, it is evident that $\dot{\sigma}_{02}$ is dependent on the diagonal elements $\sigma_{00}$, $\sigma_{11}$, and $\sigma_{22}$ in addition to other off-diagonal terms such as $\sigma_{13}$ and $\sigma_{04}$. Therefore, choosing the initial state of the particle to be a diagonal state, such as a thermal state, the vacuum will generate quantum coherence over time. In addition to the frequency shift, the generation of quantum coherence would provide a clear signature of the presence of vacuum fluctuations.

One possible way to detect the generation of these coherences is through quantum state tomography~\cite{motionalTomo1,motionalTomo2,motionalTomo3}, in which the motional state of the particle can be reconstructed through repeated measurements on identical copies of the state. This furnishes information on the full state of the system (up to a given fidelity), and therefore, the expectation value of any operator of interest. One useful example to consider is the observable $\hat{X}$ defined as
\begin{align}
    \hat{X} = \frac{ m \omega_{c}}{\hbar} \hat{x}^{2} - \frac{1}{\hbar m \omega_{c}} \hat{p}^{2} = \bop^{2} + \bdp{2}.
\end{align}
The expectation of this observable can be found from the density matrix, $\hat{\sigma}$ where $\langle \hat{X} \rangle = \text{Tr} [ \hat{\sigma} \hat{X} ]$. This relates $\langle \hat{X} \rangle$ to the elements of the density matrix
\begin{align}
    \label{eq:Xexp}
    \langle \hat{X} \rangle = \sum_{n} \left( \sqrt{n(n-1)} \sigma_{n  (n-2)} + \sqrt{(n+1)(n+2)} \sigma_{n (n+2)} \right).
\end{align}
It can be seen from Eq.~\eref{eq:Xexp} that the expectation of $\hat{X}$ is only non-zero when there are non-zero off-diagonal elements of $\hat{\sigma}$. Therefore, for an initial diagonal state of $\hat{\sigma}$, measurement of a non-zero value of $\langle \hat{X} \rangle$ provides evidence for the existence of Lines 3-5 of  Eq.~\eref{eq:mastereq2} and serves as an observable signature of vacuum fluctuations. However, this signature is not discernible given the current fidelity achievable in tomographic measurements.
The third process captured by the master equation is decay. We outline this is not a signature of vacuum fluctuations and is classical in origin.

\textit{Decay}:
The second line of Eq.~\eref{eq:mastereq2} describes damping with strength $\Gamma$. 
This damping strength in Eq.~\eqref{eq:gamma} is identical to that arising from the Larmor formula for radiation loss due to an accelerating charged particle~\cite{Jackson},
and is indistinguishable from this classical phenomenon.
The classical decay given by the Larmor formula is valid in the limit $\hbar \Omega_{\text{max}} \ll m c^{2}$, 
equivalently, $\kmax \equiv \Omega_{\text{max}}/c \ll 2 \pi/\lambda_{\text{Comp}}$ where $\lambda_{\text{Comp}}$ is the Compton wavelength~\cite{Higu2009}. This holds for the different $\wmax$ values chosen in the limit $\wmax\leqslant \sqrt{2 m c^{2} \omega_{c} /\hbar}$.

\textit{Choice of gauge}:
To assess the role played by the choice of gauge, we compare our results with the gauge-invariant results obtained by using the rotating-wave approximation (RWA) in addition to Approximations \ref{Ass:WeakField}--\ref{Ass:eqbm}. Here,
\begin{enumerate}[label=(\roman*)]
\setcounter{enumi}{5}
    \item \label{Ass:RWA} RWA implies non-resonant terms are neglected from the interaction.
\end{enumerate} 
The interaction Hamiltonian in Eq.~\eref{eq:HI} then reduces to
\begin{align}
    \hat{H}_{\textsc{i}} =  \sum_{\kb,\mu} g_{\kb,\mu} \left(\bop \ad - \bd \aop \right),
\end{align}
and the master equation becomes
\begin{align}
    \label{eq:mastereq}
    \frac{\rmd \hat{\sigma}}{\rmd t} = -  i \left( \omega_{c} + \Delta_{-} \right) \left[ \bd \bop , \hat{\sigma} \right]  + \Gamma \left( \bop \hat{\sigma} \bd - \frac{1}{2} \bd \bop \hat{\sigma} - \frac{1}{2} \hat{\sigma} \bd \bop \right).
\end{align}
This master equation is of the more familiar Lindblad form~\cite{Breuer/Petruccione}. 
Therein, approximations \ref{Ass:tscale} - \ref{Ass:RWA} ensure that it is gauge-invariant \cite{Stokes, Stokes2023}. 
Comparing Eqs.~\eref{eq:mastereq2} and \eref{eq:mastereq}, it can be seen that both the magnitude of the frequency shift and generated coherences are dependent on the gauge.
The existence of a frequency shift and decay itself are recognised as gauge-\emph{independent} processes, while  
the magnitude of the shift differs between Eqs.~\eref{eq:mastereq2} and \eref{eq:mastereq}. This is in contrast to the coherence terms of Eqs.~\eref{eq:mastereq}, where both the presence of these terms and the strength of them are strictly dependent on the gauge.

Indeed, the choice of gauge is intricately tied to the partitioning of the particle-field quantum system into subsystems, which in turn 
is determined by the operationally available interactions as well as measurements~\cite{Stokes2023}.
Physically, the RWA is typically well satisfied in quantum optical interactions. However, it is not expected to be so here
as the vacuum state of the field exists at all frequencies whose cumulative effect on the particle is sought.
We thus expect the magnitude of the frequency shift to be given by Eq.~\eqref{eq:deltanew}. The relative frequency shifts calculated for each cut-off, in the RWA limit, are displayed in Table~\ref{tab:estimates}. The present experimental sensitivity~\cite{Xing2023} is more than an order of magnitude higher than that necessary for testing the validity of the RWA. Improving the sensitivity to $10^{-11}$, there are only two possible outcomes: (i) a shift is detected validating the use of RWA, or (ii) no shifts are detected which could confirm that the RWA is not a suitable approximation in this case. The latter outcome would make the frequency shift obtained without the RWA relevant. We estimate the same in what follows.

\begin{figure}[h!]
		\includegraphics[width=0.4\textwidth]{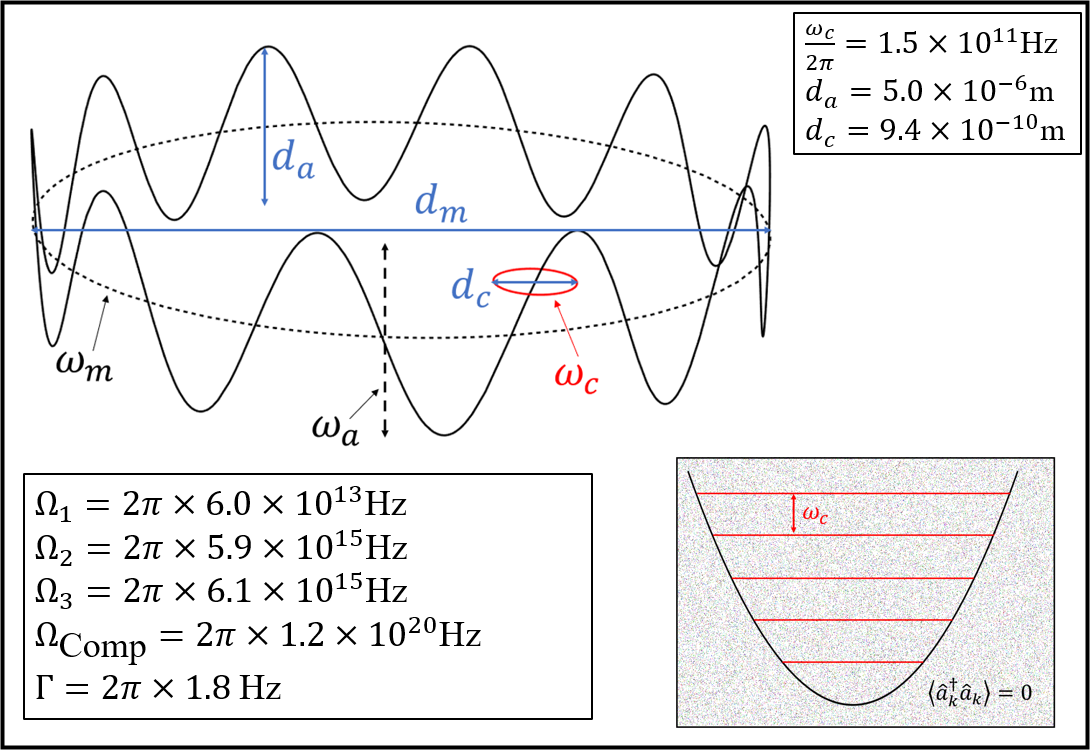}
        \caption{Motion of an electron in a Penning trap characterised by three modes of oscillation: magnetron $\omega_{m}$ (black dotted circle), axial $\omega_{a}$ (black dotted line) and cyclotron $\omega_{c}$ (red) modes with amplitudes $d_{m}$, $d_{a}$ and $d_{c}$  respectively in an EM vacuum bath $\langle \hat{a}^{\dagger}_{k} \hat{a}_{k} \rangle = 0$. Finding the frequency shift to be the largest in the cyclotron mode, we effectively model particle motion as a \textit{one-dimensional} harmonic oscillator with frequency $\omega_{c}$ (bottom-right). Adapted from \cite{Geonium}. The experimental parameters used are displayed in the top-right box. Calculated values are displayed in the bottom-left box.} 
        \label{fig:modelsketch}
\end{figure}

\textit{Experimental detection}:
We apply the frequency shift computed for a one-dimensional oscillator to an SEC~\cite{Geonium}. We show in Appendix~\ref{app:model} of the Supplementary Material that the frequency shift obtained in the context of the two-dimensional cyclotron motion of an electron depicted in Fig. \ref{fig:modelsketch}, is identical to that computed for a one-dimensional oscillator. The motion of an electron in a Penning trap consists of three separate components. In the plane of the trap, there is a slow drift magnetron motion and a fast, short-radius cyclotron motion. Perpendicular to the plane of the trap, there is a large axial oscillation on the vertical axis. We consider the shift in frequency of just one mode of motion of the cyclotron, namely, the cyclotron mode with $\omega_{c}$ as outlined in Fig. \ref{fig:modelsketch}.

The frequency, $\omega_{c}$, of the cyclotron is shifted by the presence of quantum vacuum fluctuations by $\Delta \omega$. Here, in the limit $\wmax \gg \omega_{c}$, the shift $\Delta \omega = \Delta_{-}$ (Eq.~\eref{eq:deltare}}) within the RWA, while $\Delta \omega = \Delta \omega_{c}$ (Eq.~\eref{eq:deltalimit}) beyond the RWA. As the vacuum fluctuations are always present, only the total shifted frequency, $\omega_{\textsc{t}}$, can be measured where,
\begin{align}
    \label{eq:wtotal}
    \omega_{\textsc{t}} = \omega_{c} + \Delta \omega.
\end{align}
The cyclotron frequency is linearly dependent on the magnetic field strength, $B$ as
\begin{align}
    \omega_{c} = \frac{e |B|}{m c}.
\end{align}
From Eq.~\eref{eq:deltalimit} (and respectively, Eq.~\eref{eq:deltare} within the RWA), it can be seen that the shift in the frequency has a non-linear dependence on the magnetic field. Detection of the frequency shift could therefore be possible through varying the magnetic field. The dependence on the magnetic field is given to leading order in Table~\ref{tab:Bdep} for each cut-off within and beyond the RWA. We do so for three different choices of $\wmax = \{\Omega_1,\Omega_2,\Omega_3 \}$ compatible with the LWA
for two different cases, namely, with (from Eq. \eref{eq:mastereq2}) and without (from Eq. \eref{eq:mastereq}) the RWA.

The first cut-off, $\Omega_{1}$, is obtained from assuming the wavelength, $\lambda_{k}$, of each field mode is much larger than the largest amplitude of oscillation of the particle in the trap $d_{a}$: $\lambda_{k} > d_{a}$. This places a limit on the maximum allowed frequency of the field modes, such as, 
$\Omega_{1} = 2 \pi c/d_{a}$.
The second cut-off, $\Omega_{2}$, is the frequency corresponding to the de Broglie wavelength, $\Omega_{2} = c \, m \, d_{c} \, \omega_{c}/\hbar$, of the particle.
Finally, we consider the maximum cut-off that is allowed by the LWA, i.e., 
$\Omega_{3} = \sqrt{2 m c^{2} \omega_{c}/\hbar}$.
The commonly considered cut-off $\Omega_{\text{Comp}}$ corresponding to the Compton wavelength exceeds $\Omega_{3}$ by over four orders of magnitude for the
parameters of the single-electron cyclotron.

\begin{table}[h]
\begin{tabular}{|c | c | c | c |} 
 \hline
 $\Delta \omega(B)$ & $\rule{0cm}{0.5cm}\Omega_{1}$ \Equal $\frac{2 \pi c}{d_{a}}$ & $\Omega_{2}$ \Equal $\frac{c  m  d_{c}  \omega_{c}}{\hbar} $ & $\Omega_{3}$ \Equal $\sqrt{\frac{2 m c^{2} \omega_{c}}{\hbar}}$ \\[5pt]
  \hline
 With RWA & $\rule{0cm}{0.5cm} \sim |B|^{2} \ln{|B|}$ & $\sim |B|^{2}$ & $\sim |B|^{2} \ln{|\sqrt{B}|}$ \\[5pt]
 \hline
 Without RWA & $\rule{0cm}{0.5cm} \sim |B|^{3}$ & $\sim |B|^{2}$ & $\sim |B|^{\frac{5}{2}}$ \\[5pt]
 \hline
\end{tabular}
\caption{The dependence of the cyclotron frequency shift, $\Delta \omega$, on the magnetic field of the SEC is given to leading order for each cut-off $\Omega_{1}$, $\Omega_{2}$, $\Omega_{3}$ with and without the RWA. This is in the limit $\wmax \gg \omega_{c}$.} 
\label{tab:Bdep}
\end{table}

It can be seen from Table~\ref{tab:Bdep} that the shift in the cyclotron frequency depends on whether the RWA is applied and on the choice of the frequency cut-off. It is interesting to note that the dependence on the magnetic field within the RWA, is approximately $B^{2}$ for all choices of cut-off (up to some logarithmic factor). This is in contrast to the cases beyond the RWA, where the magnetic field dependence in the frequency shift is specific to the choice of cut-off.
In addition to confirming the presence of vacuum fluctuations, measurement of the exact magnetic field dependence of $\Delta \omega$ could also give evidence to the correct choice of cut-off, specifically if shifts obtained beyond the RWA become relevant.

\textit{Conclusion}:
We have identified two signatures of the quantum vacuum fluctuations of the EM field: a frequency shift Eq.~\eref{eq:deltanew} and generation of coherences Eq.~\eref{eq:sigma02}. 
Our results are gauge-dependent as our analysis goes beyond the RWA, and cut-off dependent - consistent with the LWA.
Our estimates using parameters of modern single-electron cyclotrons suggest that these shifts and coherences should be detectable in future experiments.
That will be the ultimate adjudicator of the validity of our approximations and analysis. We must however note that a more accurate study of the SEC would require the inclusion of discrete cavity modes and the magnetic interaction of the electron with the vacuum.

\textit{Acknowledgements:} 
We thank the two anonymous referees for improving this work considerably.
We thank Peter Barker, Xing Fan and Frank Schlawin for fruitful discussions. This work has been funded, in part, by the UK STFC “Quantum Technologies for Fundamental Physics” program (Grant Numbers ST/Y004493/1, ST/T006404/1 and ST/W006308/1).

\bibliography{refs}

\onecolumngrid

\onecolumngrid
\appendix
\textbf{\large\underline{Supplementary Material}}

\section{\label{app:spinint} Magnitude of Spin Interaction Term}

The particle's spin will also interact with the vacuum state of the EM field through a $\mu \cdot \mathbf{B}$ interaction. Where $\mu$ is the magnetic moment moment of the particle and $B$ is the magnetic field. The magnetic moment is related to the particle's spin, $S=\left(\frac{1}{2}, -\frac{1}{2}\right)$ through the expression
\begin{align}
    \mu = \frac{g q \hbar S}{4 m c},
\end{align}
where the g-value is $g\approx2$.
The magnetic field, $\mathbf{B}$, can be quantized in a similar approach to that used in obtaining Eq. \eref{eq:Afield}
\begin{align}
    \mathbf{B}(\mathbf{r},t) = \frac{i}{c}\sum_{\mathbf{k}, \mu} \left(\frac{\mathbf{k}}{|\mathbf{k}|} \times \mathbf{s}_{\mathbf{k}, \mu} \right) \sqrt{\frac{\hbar \wk}{2 \epsilon_{0} V}} \left(\hat{a}_{\mathbf{k}}(t) e^{i \mathbf{k} \cdot \mathbf{r}} - \hat{a}^{\dagger}_{\mathbf{k}}(t) e^{-i \mathbf{k} \cdot \mathbf{r}}  \right)
\end{align} 
The Hamiltonian interaction of Eq. \eref{eq:Hfi} would then be extended to
\begin{align}
    \hat{H}_{\textsc{i}} = \sum_{\kb,\mu} g_{\kb,\mu} \left(\bop - \bd \right)\left( \aop + \ad \right) + \sum_{\kb,\mu} h_{\kb,\mu} (\aop - \ad),
\end{align}
where the LWA has been used once again and constants $g_{\kb,\mu}$ and $h_{\kb,\mu}$ are
\begin{align}
    \label{eq:gk2}
    g_{\kb,\mu} &=  \frac{q \hbar}{2 i } \, \sqrt{\frac{\omega_{c}}{\epsilon_{0} m \Omega_{\mathbf{k}} V}} \, \cos \theta_{\kb,\mu}, \\
    h_{\kb,\mu} &= \frac{i g q \hbar S}{4 mc^{2}} \sqrt{\frac{\hbar \wk}{2 \epsilon_{0} V}}\cos \theta'_{\kb,\mu},
\end{align}
where $\theta_{\kb,\mu}$ is the angle between $\mathbf{s}_{\mathbf{k}, \mu}$ and the axis along which the particle motion is confined, namely, $\mathbf{e}_{z}$ and $\theta'_{\kb,\mu}$ is the angle between $\mathbf{k} \times \mathbf{s}_{\mathbf{k}, \mu}$ and $\mathbf{e}_{z}$.

An order of magnitude comparison between the two couplings $g_{\kb,\mu}$ and $h_{\kb,\mu}$ is then made
\begin{align}
    \left|\frac{g_{\kb,\mu}}{h_{\kb,\mu}}\right| \approx \frac{c^{2}}{\wk} \sqrt{\frac{m \omega_{c}}{\hbar}}.
\end{align}
For the parameter values defined in Fig. \ref{fig:modelsketch} this becomes
\begin{align}
    \left|\frac{g_{\kb,\mu}}{h_{\kb,\mu}}\right| \approx \frac{10^{24}s^{-1}}{\wk}.
\end{align}
Therefore, so long as $\wk \ll 10^{24}s^{-1}$, it is safe to neglect the interaction of the particle's spin. The greatest cut-off we consider is $\Omega_{3} \approx 10^{16}$Hz. We are therefore well within this bound for the experiment considered.

\section{\label{app:lwa} Validity of the Long-Wavelength Approximation}

The quantisation of the EM field Eq.\eref{eq:Afield} contained the exponential $e^{i \kb \cdot \mathbf{r}}$. The series expansion of which is
\begin{align}
    e^{i \kb \cdot \mathbf{r}} = 1 + i \kb \cdot \mathbf{r} - \frac{\left( \kb \cdot \mathbf{r} \right)^{2}}{2} + ...
\end{align}
The interaction Eq.\eref{eq:HI} up to second order in $\kb \cdot \mathbf{r}$ is 
\begin{align}
    H_{\textsc{i}} = \sum_{\mathbf{k}} g'_{\mathbf{k}} \: p \left( \left( \aop + \ad \right) + i k \: r \cos{\phi_{\kb}} \left( \aop - \ad \right) - \frac{(k \: r)^{2}}{2} \cos^{2}{\phi_{\kb}} \left( \aop + \ad \right) \right)
\end{align}
where
\begin{align}
    g'_{\mathbf{k}} = \sum_{\kb} \frac{q}{m} \sqrt{\frac{\hbar}{2 \wk \epsilon_{0} V}}.
\end{align}
Symmeterizing in $r$ and $p$
\begin{align}
     H_{\textsc{i}} = \sum_{\mathbf{k}} g'_{\mathbf{k}} \left( p \left( \aop + \ad \right) + \frac{i k}{2} \cos{\phi_{\kb}}\: \left(p r +r p \right) \left( \aop - \ad \right) - \frac{k^{2}}{6} \cos^{2}{\phi_{\kb}} \left(r^{2} p + rpr + p r^{2} \right) \left( \aop + \ad \right) \right).
\end{align}
The Hamiltonian is quantized by replacing $r, p$ with $\hat{x}$ and ${p}$ where
\begin{align}
    \hat{x} &= \sqrt{\frac{\hbar}{2 m \omega_{c}}} \left(\bop + \bd \right), \\
    \hat{p} &= -i\sqrt{\frac{m \omega_{c} \hbar}{2}} \left(\bop - \bd \right).
\end{align}
The resulting Hamiltonian is
\begin{align}
     H_{\textsc{i}} = \sum_{\mathbf{k}} \left( g^{(0)}_{\kb} \left( \bop - \bd \right) \left( \aop + \ad \right) + g^{(1)}_{\kb} \left(\bop^{2} - \bdp{2} \right) \left( \aop - \ad \right) +g^{(2)}_{\kb} \left(\bop^{3} - \bdp{3} +  \bop^{2} \bd -  \bop \bdp{2} +  \bd -  \bop \right) \left( \aop + \ad \right) \right)
\end{align}
where
\begin{align}
    g_{\kb}^{(0)} &= \frac{q \hbar}{ i} \, \sqrt{\frac{\omega_{c}}{\epsilon_{0} m \Omega_{\mathbf{k}} V}} \, \cos \theta_{\kb,\mu}, \\
    g_{\kb}^{(1)} &= \frac{q}{m c} \sqrt{\frac{\hbar^{3} \wk}{2 \epsilon_{0} V}}\cos \theta_{\kb,\mu} \cos{\phi_{\kb}}, \\
    g_{\kb}^{(2)} &= \frac{q \hbar^{2}}{4 i c^{2}} \sqrt{\frac{\wk^{3}}{\omega_{c} \epsilon_{0} m^{3} V}} \cos \theta_{\kb,\mu} \cos^{2}{\phi_{\kb}},
\end{align}
and a factor of 2 has been included to sum over the two independent polarization states.
The angular dependence of $\theta_{\kb,\mu}$ and $\phi_{\kb}$ is neglected as it is at most order 1. The constants are then
\begin{align}
    g_{\kb}^{(0)} &= \frac{q \hbar}{ i} \, \sqrt{\frac{\omega_{c}}{\epsilon_{0} m \Omega_{\mathbf{k}} V}}, \\
    g_{\kb}^{(1)} &= \frac{q}{m c} \sqrt{\frac{\hbar^{3} \wk}{2 \epsilon_{0} V}}, \\
    g_{\kb}^{(2)} &= \frac{q \hbar^{2}}{4 i c^{2}} \sqrt{\frac{\wk^{3}}{\omega_{c} \epsilon_{0} m^{3} V}}.
\end{align}
The shift to the energy levels of the oscillator can be calculated using second-order perturbation theory
\begin{align}
    \Delta E_{n} &= \sum_{m \neq n, k_{l} \neq 0 } \frac{|\bra{m \otimes k_{l}} H_{\textsc{i}} \ket{n \otimes 0} |^{2}}{E^{(0)}_{n, 0} - E^{(0)}_{m, k_{l}}} \nonumber \\
    &= \sum_{m \neq n, k_{l} \neq 0 } \frac{|\bra{m \otimes k_{l}} V^{(0)} + V^{(1)} + V^{(2)} \ket{n \otimes 0} |^{2}}{E^{(0)}_{n, 0} - E^{(0)}_{m, k_{l}}}.
\end{align}
where
\begin{align}
    V^{(0)} &= \sum_{\kb} g_{\kb}^{(0)} \left( \bop - \bd \right) \left( \aop + \ad \right), \\
    V^{(1)} &= \sum_{\kb} g_{\kb}^{(1)} \left(\bop^{2} - \bdp{2} \right) \left( \aop - \ad \right), \\
     V^{(2)} &= \sum_{\kb} g_{\kb}^{(2)}  \left(\bop^{3} - \bdp{3} +  \bop^{2} \bd -  \bop \bdp{2} +  \bd -  \bop \right) \left( \aop + \ad \right).
\end{align}
Calculating the expectation in the numerator
\begin{align}
    \bra{m \otimes k_{l}} V^{(0)} \ket{n \otimes 0} = \sum_{\kb} g^{(0)} \Big( \delta_{m,n-1} \delta_{k_{l}, 1} \sqrt{n}  \nonumber - \delta_{m,n+1} \delta_{k_{l}, 1} \sqrt{n+1} \Big),
\end{align}
\begin{align}
    \bra{m \otimes k_{l}} V^{(1)} \ket{n \otimes 0} = \sum_{\kb} g^{(1)} &\Big(-  \delta_{m,n-2} \delta_{k_{l},1} \sqrt{n(n-1)} + \delta_{m,n+2} \delta_{k_{l},  1} \sqrt{(n+1)(n+2)} \Big),
\end{align}
\begin{align}
    \bra{m \otimes k_{l}} V^{(2)} \ket{n \otimes 0} = \sum_{\kb} g^{(2)} &\Big( \delta_{m,n-3} \delta_{k_{l}, 1} \sqrt{n(n-1)(n-2)} - \delta_{m,n+3} \delta_{k_{l}, 1} \sqrt{(n+1)(n+2)(n+3)}  \nonumber \\
    &+\delta_{m,n-1} \delta_{k_{l}, 1} (n+1)\sqrt{n} - \delta_{m,n+1} \delta_{k_{l}, 1} (n+2)\sqrt{n+1}  \nonumber \\
    &+\delta_{m,n+1} \delta_{k_{l}, 1} \sqrt{n+1} - \delta_{m,n-1} \delta_{k_{l}, 1} \sqrt{n}  \Big),
\end{align}
\begin{align}
    |\bra{m \otimes k_{l}} V^{(0)} + V^{(1)} + V^{(2)} \ket{n \otimes 0} |^{2} = |g^{(0)}|^{2} \delta_{k_{l},1} &\Big( \delta_{m,n-1} n + \delta_{m,n+1} (n+1) \Big) \nonumber \\
    +|g^{(1)}|^{2} \delta_{k_{l},1} &\Big( \delta_{m,n-2} n(n-1) + \delta_{m,n+2} (n+1)(n+2) \Big) \nonumber \\
    +|g^{(2)}|^{2} \delta_{k_{l},1} &\Big( \delta_{m,n-3} n (n-1) (n-2) + \delta_{m,n+3} (n+1)(n+2)(n+3) \nonumber \\
    &+ \delta_{m,n-1}  n^{3}  + \delta_{m,n+1}  (n+1)^{3} \Big) \nonumber \\
    +2|g^{(0)} g^{(2)}| \delta_{k_{l},1} &\Big( \delta_{m,n-1}  n^{2} + \delta_{m,n+1}  (n^{2} + 2n +1) \Big).
\end{align}
The resulting energy shift is
\begin{align}
    \Delta E_{n} &= \sum_{k} \frac{|g^{(0)}|^{2} n}{\hbar \omega_{c} - \hbar \wk} - \sum_{k} \frac{|g^{(0)}|^{2} (n+1)}{\hbar \omega_{c} + \hbar \wk} + \sum_{k} \frac{|g^{(1)}|^{2} n(n-1)}{2\hbar \omega_{c} - \hbar \wk} - \sum_{k} \frac{|g^{(1)}|^{2} (n+1)(n+2)}{2 \hbar \omega_{c} + \hbar \wk} \nonumber \\
    &+ \sum_{k} \frac{|g^{(2)}|^{2}  n(n-1)(n-2)}{3\hbar \omega_{c} - \hbar \wk} - \sum_{k} \frac{|g^{(2)}|^{2}  (n+1)(n+2)(n+3)}{3\hbar \omega_{c} + \hbar \wk} \nonumber \\
    &+ \sum_{k} \frac{|g^{(2)}|^{2} n^{3}}{\hbar \omega_{c} - \hbar \wk} - \sum_{k} \frac{|g^{(2)}|^{2} (n^{3} + 3n^{2} + 3 n - 1)}{\hbar \omega_{c} + \hbar \wk} \nonumber \\
    &+ \sum_{k} \frac{2 |g^{(0)}g^{(2)}|  n^{2}}{\hbar \omega_{c} - \hbar \wk} - \sum_{k} \frac{2 |g^{(0)}g^{(2)}|  (n^{2} + 2n +1)}{\hbar \omega_{c} + \hbar \wk}.
\end{align}
This can be rewritten in terms of the following constants
\begin{align}
    \frac{\Delta E_{n}}{\hbar} &= n \Delta^{(0)}_{-} - (n+1) \Delta^{(0)}_{+} + n(n-1) \Delta^{(1)}_{-} - (n+1)(n+2) \Delta^{(1)}_{+} +  n(n-1)(n-2) \Delta^{(2a)}_{-} \nonumber \\
    &-  (n+1)(n+2)(n+3) \Delta^{(2a)}_{+} + n^{3} \Delta^{(2b)}_{-} - (n^{3} +3n^{2} + 3n-1) \Delta^{(2b)}_{+} \nonumber \\
    &+ n^{2} \Delta^{(2c)}_{-} - (n^{2} + 2n +1) \Delta^{(2c)}_{+}
    \label{eq:freqShift}
\end{align}
where
\begin{align}
    \Delta_{\pm}^{(0)} &= \frac{1}{\hbar}\sum_{k} \frac{|g^{(0)}|^{2}}{\hbar \omega_{c} \pm \hbar \wk} = \frac{2 \alpha_{\textsc{fs}} \kappa}{ \pi} \left(- \omega_{c} \ln \left|\frac{\omega_{c} \pm \wmax}{\omega_{c}}\right| \pm \wmax \right), \label{eq:delta0} \\
    \Delta_{\pm}^{(1)} &= \frac{1}{\hbar} \sum_{k} \frac{|g^{(1)}|^{2}}{2\hbar \omega_{c} \pm \hbar \wk} = \frac{\alpha_{\textsc{fs}} \kappa^{2}}{\pi }\left(- 8 \omega_{c} \ln \left|\frac{2\omega_{c} \pm \wmax}{2 \omega_{c}} \right| \pm 4 \wmax - \frac{\wmax^{2}}{\omega_{c}} \pm \frac{\wmax^{3}}{3 \omega_{c}^{2}}\right) \\
    \Delta_{\pm}^{(2a)} &= \frac{1}{\hbar} \sum_{k} \frac{|g^{(2)}|^{2}}{3\hbar \omega_{c} \pm \hbar \wk} = \frac{\alpha_{\textsc{fs}} \kappa^{3}}{8 \pi } \left(-243 \omega_{c} \ln \left|\frac{3\omega_{c} \pm \wmax}{3\omega_{c}} \right| \pm 81 \wmax - \frac{27 \wmax^{2}}{2 \omega_{c}} \pm \frac{3  \wmax^{3}}{\omega_{c}^{2}} - \frac{3 \wmax^{4}}{4\omega_{c}^{3}} \pm \frac{\wmax^{5}}{5\omega_{c}^{4}} \right) \nonumber \\
    \Delta_{\pm}^{(2b)} &= \frac{1}{\hbar} \sum_{k} \frac{|g^{(2)}|^{2}}{\hbar \omega_{c} \pm \hbar \wk} = \frac{\alpha_{\textsc{fs}} \kappa^{3}}{ 8 \pi } \left(-\omega_{c} \ln \left|\frac{\omega_{c} \pm \wmax}{\omega_{c}} \right| \pm \wmax - \frac{ \wmax^{2}}{2 \omega_{c}} \pm \frac{ \wmax^{3}}{3 \omega_{c}^{2}} - \frac{ \wmax^{4}}{4 \omega_{c}^{3}} \pm \frac{\wmax^{5}}{5\omega_{c}^{4}} \right) \nonumber , \\
    \Delta_{\pm}^{(2c)} &= \frac{1}{\hbar} \sum_{k} \frac{2 |g^{(0)} g^{(2)}|}{\hbar \omega_{c} \pm \hbar \wk} =  \frac{\alpha_{\textsc{fs}} \kappa^{2}}{\pi } \left(-\omega_{c} \ln \left|\frac{\omega_{c} \pm \wmax}{\omega_{c}} \right| \pm  \wmax - \frac{ \wmax^{2}}{2\omega_{c}} \pm \frac{\wmax^{3}}{3\omega_{c}^{2}} \right),
\end{align}
and $q$ is now taken as the electron charge $e$ and $\kappa = \hbar \omega_{c} / mc^{2}$.

To obtain the true frequency shift of the harmonically-trapped particle, we need to renormalize away a contribution arising from the free particle interaction. However, we are currently only interested in checking the validity of the long-wavelength approximation. So we do not resort to a full renormalization. Instead, we first obtain the limit at which the long-wavelength approximation is valid. We then obtain the true shift in the presence of the long-wavelength approximation, or equivalently, only renormalize $\Delta_{\pm}^{(0)}$ with the obtained limit. 

For the long-wavelength approximation to be valid, we need $|\Delta_{\pm}^{(0)}|$ should be greater than $|\Delta_{\pm}^{(1)}|$, $|\Delta_{\pm}^{(2a)}|$, $|\Delta_{\pm}^{(2b)}|$, and $|\Delta_{\pm}^{(2c)}|$. Comparing the absolute value of the largest contributions to each of these terms up to an order of magnitude, we find
\begin{align}
    \kappa \wmax > \kappa^{2} \frac{\wmax^{3}}{\omega_{c}^{2}} \implies \wmax < \sqrt{\frac{m c^{2} \omega_{c}}{\hbar}}.
\end{align}
This is not too far from the limit that would be found by considering
\begin{align}
    e^{kx} \approx 1 &\implies \kmax x < 1, \nonumber \\
    &= \frac{\wmax}{c} \sqrt{\frac{\hbar}{2m\omega_{c}}} < 1 \nonumber \\
    &= \wmax < \sqrt{\frac{2 m c^{2} \omega_{c}}{\hbar}}.
\end{align}
For the single-electron cyclotron considered, the long-wavelength approximation is valid so long as 
\cite{Gabrielse}, $\wmax /2\pi < 6.1 \times 10^{15}$ Hz.

Therefore, assuming the validity of the long-wavelength approximation, we find the renormalized frequency shift using Eqs. \eref{eq:freqShift} and \eref{eq:delta0} to compare with that obtained using the master equation approach. Considering the limit $\omega_{c}\to 0$ in Eq.\eref{eq:delta0}, it can be seen that the term $\pm 2 \alpha_{\textsc{fs}} \kappa \wmax/\pi$ is the free particle energy shift. This term is dropped in Eq. \eref{eq:delta0} to obtain the frequency shift using Eq. \eref{eq:freqShift}, as
\begin{align}
    \Delta \omega_{c} = -\frac{2 \alpha_{\textsc{fs}} \kappa \omega_{c}}{ \pi } \ln \left|\frac{\omega_{c} - \wmax}{\omega_{c} + \wmax}\right| \approx -\frac{4 \alpha_{\textsc{fs}} \kappa \omega_{c}^{2}}{ \pi } \frac{1}{\wmax}.
\end{align}
This is in agreement with the result obtained from our master equation approach. We point out that any differences up to O(1) arise from neglecting factors such as $\cos \theta_{\kb,\mu}$.

\section{\label{app:MEderiv} Deriving the Master Equation}

In this section, we discuss the salient steps involved in obtaining the master equation \eref{eq:mastereq} and \eref{eq:mastereq2}. 
We recall Eq. \eref{eq:HI} that 
\begin{align}
\label{eq:HIapp}
\hat{H}_{\textsc{i}} =  \sum_{\kb,\mu} g_{\kb,\mu} \left(\bop - \bd \right) \left( \aop  + \ad \right),
\end{align}
with the coupling strength (Eq. \eref{eq:gk})
\begin{align*}
    g_{\kb,\mu} =  \frac{q \hbar}{2 i } \, \sqrt{\frac{\omega_{c}}{\epsilon_{0} m \Omega_{\mathbf{k}} V}} \, ( \mathbf{s}_{\mathbf{k}, \mu} \cdot \mathbf{e}_{x} ).
\end{align*}
Here $\epsilon_{0}$ is the permittivity of free space, $V$ is an arbitrary normalisation volume. For each mode, there will be two independent polarisations $\mu\in \{+,\times\}$. As the contributions from either polarisation are identical, we account for such contributions by
\begin{align}
    \label{eq:gkApp}
    g_{\kb} = 2 \: g_{\kb,\mu} =  \frac{q \hbar}{ i } \, \sqrt{\frac{\omega_{c}}{\epsilon_{0} m \Omega_{\mathbf{k}} V}} \, ( \mathbf{s}_{\mathbf{k}, \mu} \cdot \mathbf{e}_{x} ).
\end{align}

We want to study the dynamics of the particle subsystem using its time-evolved reduced density matrix $\hat{\sigma}(t) = \text{Tr}_{\textsc{f}}[ \hat{\rho}(t) ]$ where $\hat{\rho}(t)$ denotes the density matrix of the full particle-field system. The density matrix corresponding to the field $\hat{\sigma}_{\textsc{f}} (t) = \text{Tr}_{\textsc{p}}[ \hat{\rho}(t) ]$. We recall that the state of the field is assumed to be \textit{stationary}, i.e., we assume $\hat{\sigma}_{\textsc{f}} (t)$ is independent of $t$.
We also recall that we work in the interaction picture such that
\begin{subequations}
\begin{align}
    \bopt (t) &= \bop(0) e^{- i \omega_{c} t} \equiv \bop e^{-i \omega_{c} t}, \\
    \aopt (t) &= \aop(0) e^{- i \Omega_{\kb} t} \equiv \aop e^{-i \wk t},
\end{align}
\label{eq:shiftOps}
\end{subequations}
obtained using
\begin{align}
    \widetilde{\mathcal{O}}(t) = e^{\frac{i}{\hbar}\left(\hat{H}_{P} + \hat{H}_{\textsc{f}} \right)t} \hat{\mathcal{O}} (0) e^{-\frac{i}{\hbar}\left(\hat{H}_{P} + \hat{H}_{\textsc{f}} \right)t},
\end{align}
for any operator $\hat{\mathcal{O}}$. We also use the notation $\hat{\mathcal{O}}(0) = \hat{\mathcal{O}}$.
Using the assumptions \ref{Ass:weakInt}, \ref{Ass:eqbm} and \ref{Ass:tscale} the Liouville equation to second-order in the interaction strength is presented~\cite{Cohen}, 
\begin{align}
    \label{eq:evoeqa1}
    \frac{\Delta \widetilde{\sigma}(t)}{\Delta t} = -\frac{1}{\hbar^{2}} \frac{1}{\Delta t} \int^{\infty}_{0} d(t_{1} - t_{2}) \int_{t}^{t + \Delta t} d t_{1} \text{Tr}_{\textsc{f}} [ \widetilde{H}_{\textsc{i}} ( t_{1} ), [ \widetilde{H}_{\textsc{i}} ( t_{2} ) , \widetilde{\sigma} (t) \otimes \widetilde{\sigma}_{\textsc{f}} ]].
\end{align}
We note here that the contributions from the second-order terms include those that are dependent on the commutator of the field operators. This is crucial in obtaining signatures that are specific to the quantum nature of the field.

Partial trace over the field subsystem with terms having $(\aopt)^{2} \sigmat_{\textsc{f}}$ and $(\adt)^{2} \sigmat_{\textsc{f}}$ vanish. This is by virtue of Assumption \ref{Ass:eqbm} as it requires the field to be a statistical mixture of the eigenstates of $H_{\textsc{f}}$. Using this in addition to Eqs. \eref{eq:HIapp}, \eref{eq:shiftOps} and \eref{eq:evoeqa1}, we obtain
\begin{align}
    \label{eq:evoeqa2}
    \frac{\Delta \widetilde{\sigma}(t)}{\Delta t} &= -\frac{1}{\hbar^{2}} \frac{1}{\Delta t} \int^{\infty}_{0} d(t_{1} - t_{2}) \int_{t}^{t + \Delta t} d t_{1} \Big( \nonumber \\
    &\text{Tr}_{\textsc{f}} \left[ \left( \bopt e^{-i \omega_{c} t_{1}} - \bdt e^{i \omega_{c} t_{1}} \right) R(t_{1}) , \left[ \left( \bdt e^{i \omega_{c} t_{2}} - \bopt e^{-i \omega_{c} t_{2}} \right) R^{\dagger} (t_{2}), \widetilde{\sigma} (t) \otimes \widetilde{\sigma}_{\textsc{f}} \right]\right] \Big) + \text{h.c,}
\end{align}
where,
\begin{align}
    R (t) &=  \sum_{\kb} g_{\kb} \aopt e^{-i \wk t}.
\end{align}
Evaluating the double commutator in Eq. \eref{eq:evoeqa2} we obtain
\begin{align}
\label{eq:evoeqa3}
\frac{\Delta \widetilde{\sigma}}{\Delta t} = -\frac{1}{h^{2}} \frac{1}{\Delta t} \int^{\infty}_{0} d(\tone - \ttwo) \int_{t}^{t + \Delta t} d \tone & \bigg(  (\bopt \bdt \sigmat - \bdt \sigmat \bopt ) \phi^{*}(t_{1} - t_{2})  \nonumber \\
- &   ( \bopt \bopt \sigmat - \bopt \sigmat \bopt) \phi^{*}(t_{1}+t_{2})   \nonumber \\
- &  ( \bdt \bdt \sigmat - \bdt \sigmat \bdt ) \phi(t_{1}+t_{2})  \nonumber \\
+ &  ( \bdt \bopt \sigmat - \bopt \sigmat \bdt  ) \phi(t_{1}-t_{2})  \bigg) \text{Tr}_{\textsc{f}} \left[ R(t_{1}) R^{\dagger} (t_{2}) \sigmat_{\textsc{f}} \right] \nonumber \\
+ & \bigg(  -( \bopt \sigmat \bdt - \sigmat \bdt \bopt ) \phi^{*}(t_{1}-t_{2})   \nonumber \\
+ &  ( \bopt \sigmat \bopt - \sigmat \bopt \bopt ) \phi^{*}(t_{1}+t_{2})   \nonumber \\
+ &  ( \bdt \sigmat \bdt - \sigmat \bdt \bdt ) \phi(t_{1}+t_{2}) \nonumber \\
- &  ( \bdt \sigmat \bopt - \sigmat \bopt \bdt ) \phi(t_{1}-t_{2})   \bigg)  \text{Tr}_{\textsc{f}} \left[ R^{\dagger} (t_{2}) R(t_{1}) \sigmat_{\textsc{f}} \right] \nonumber \\
+ & h.c,
\end{align}
where $\phi(t) = e^{i \omega_{c} t}$.
The partial trace over the field then gives
\begin{subequations}
\begin{align}
\label{eq:trace1}
\text{Tr}_{\textsc{f}} \left[ R(t_{1}) R^{\dagger} (t_{2}) \sigmat_{\textsc{f}} \right] &=  \sum_{\kb} |g_{\kb}|^{2} ( \langle n_{\kb} \rangle + \langle [\aop, \ad] \rangle ) e^{  - i \wk ( \tone - \ttwo ) }, \\
\label{eq:trace2}
\text{Tr}_{\textsc{f}} \left[ R^{\dagger} (t_{2}) R(t_{1}) \sigmat_{\textsc{f}} \right] &= \sum_{\kb} |g_{\kb}|^{2} \langle n_{\kb} \rangle  e^{ i \wk ( \tone - \ttwo )  },
\end{align}
\label{eq:traces}
\end{subequations}
where $\langle n_{\kb} \rangle \equiv \langle \ad \aop \rangle$ for each mode $\kb$. We retain the form of the term $\langle [\aop, \ad] \rangle$ (which is evidently equal to 1) to show the quantum origin of certain aspects of the dynamics, such as the level shifts and damping strength.

Defining $\tau = t_{1} - t_{2}$ and using Eq. \eref{eq:traces}, Eq. \eref{eq:evoeqa3} can now be rewritten as
\begin{align}
\label{eq:evoeqa4}
\frac{\Delta \widetilde{\sigma}}{\Delta t} = -\frac{1}{h^{2}}  \frac{1}{\Delta t} \sum_{\kb} |g_{\kb}|^{2} \int^{\infty}_{0} d \tau \int_{t}^{t + \Delta t} d \tone & \bigg( ( \bopt \bdt \sigmat - \bdt \sigmat \bopt ) \Phi_{+} (\tau)   \nonumber \\
- &  ( \bopt^{2} \sigmat - \bopt \sigmat \bopt  ) \Phi_{-} (\tau) \left( \phi^{*}(t_{1}) \right)^{2}  \nonumber \\
- &  ( \bdtp{2} \sigmat - \bdt \sigmat \bdt  ) \Phi_{+} (\tau) \left( \phi(t_{1}) \right)^{2}  \nonumber \\
+ & (\bdt \bopt \sigmat - \bopt \sigmat \bdt ) \Phi_{-} (\tau)   \bigg) ( \langle n_{\kb} \rangle + \langle [\aop, \ad] \rangle ) \nonumber \\
+ & \bigg( - ( \bopt \sigmat \bdt - \sigma \bdt \bopt ) \Phi_{+} (\tau)    \nonumber \\
+ & ( \bopt \sigmat \bopt - \sigmat \bopt^{2} ) \Phi_{-} (\tau) \left( \phi^{*}(t_{1}) \right)^{2}   \nonumber \\
+ & ( \bdt \sigmat \bdt - \sigmat \bdtp{2}  ) \Phi_{+} (\tau) \left( \phi(t_{1}) \right)^{2}  \nonumber \\
- &  ( \bdt \sigmat \bopt - \sigmat \bopt \bdt  ) \Phi_{-} (\tau) \bigg)  \langle n_{\kb} \rangle \nonumber \\
+ & h.c
\end{align}
where $\Phi_{\pm} (\tau) = e^{- i \tau \left( \wk \pm \omega_{c} \right)}$.

The integral over $\tau$ can be found using
\begin{align}
\label{eq:ints}
\int\limits_{0}^{\infty}d\tau e^{i \tau (\omega_{c} \pm \wk)} &= \frac{i}{\omega_{c} \pm \wk} + \pi \delta (\wk \pm \omega_{c}).
\end{align}
Integrating the master equation over $\tau$ and $t_{1}$ results in
\begin{align}
\frac{\Delta \widetilde{\sigma}}{\Delta t} =  & \left( \bdt \sigmat \bopt - \bopt \bdt \sigmat - \zeta(t)  ( \bdt \sigmat \bdt - \bdtp{2} \sigmat ) \right) \left( \frac{1}{2}\Gamma'_{+} + \frac{1}{2}\Gamma_{+}  - i\Delta_{+} - i\Delta'_{+} \right)   \nonumber \\
+ &  \left( \bopt \sigmat \bdt - \bdt \bopt \sigmat - \zeta^{*}(t)  ( \bopt \sigmat \bopt - \bopt^{2} \sigmat ) \right) \left( \frac{1}{2}\Gamma'_{-} + \frac{1}{2}\Gamma_{-}  + i\Delta_{-} + i\Delta'_{-} \right) \nonumber \\
+ &  \left(\bopt \sigmat \bdt - \sigmat \bdt \bopt - \zeta(t)   ( \bdt \sigmat \bdt - \sigmat \bdtp{2}  ) \right) \left( \frac{1}{2}\Gamma'_{+} - i\Delta'_{+} \right) \nonumber \\
+ &  \left( \bdt \sigmat \bopt - \sigmat \bopt \bdt  - \zeta^{*}(t)  ( \bopt \sigmat \bopt - \sigmat \bopt^{2} ) \right) \left( \frac{1}{2}\Gamma'_{-} + i\Delta'_{-} \right) + h.c,
\end{align}
where, 
\begin{subequations}
\begin{align}
\label{eq:constants1}
\Gamma_{\pm} &= \sum_{\kb} \frac{|g_{\kb}|^{2}}{\hbar^{2}} 2 \pi \, \delta \left( \wk \pm \omega_{c} \right) \: \langle [\aop, \ad] \rangle, \\
\label{eq:constants2}
\Gamma_{\pm}^{'} &= \sum_{\kb} \frac{|g_{\kb}|^{2}}{\hbar^{2}} 2 \pi \, \delta \left( \wk \pm \omega_{c} \right) \:  \langle n_{k} \rangle,  \\
\label{eq:constants3}
\Delta_{\pm} &= \sum_{\kb} \frac{|g_{\kb}|^{2}}{\hbar^{2}} \frac{1}{\omega_{c} \pm \wk} \langle [\aop, \ad] \rangle,  \\
\label{eq:constants4}
\Delta_{\pm}^{'} &= \sum_{\kb} \frac{|g_{\kb}|^{2}}{\hbar^{2}} \frac{1}{\omega_{c} \pm \wk} \langle n_{k} \rangle.
\end{align} 
\label{eqn:constset}
\end{subequations}
and defining the following phase factor $\zeta$
\begin{align}
    \zeta(t) = \int_{t}^{t + \Delta t} d t_{1} \; \left[\phi(t_{1})\right]^{2} = \frac{e^{i 2 \omega_{c} t}( e^{2 i \omega_{c} \Delta t} - 1 )}{2 i \omega_{c} \Delta t}.
    \label{eq:zetaDefn}
\end{align}
Noting that in the limit $\Delta t \rightarrow 0, \zeta(t) \rightarrow e^{2i \omega_{c} t} \equiv \phi(2 t)$, the master equation in this limit becomes
\begin{align}
\frac{\rmd \widetilde{\sigma}}{\rmd t} =  & \left( \bdt \sigmat \bopt - \bopt \bdt \sigmat - \phi(2 t)  ( \bdt \sigmat \bdt - \bdtp{2} \sigmat ) \right) \left( \frac{1}{2}\Gamma'_{+} + \frac{1}{2}\Gamma_{+}  - i\Delta_{+} - i\Delta'_{+} \right)   \nonumber \\
+ &  \left( \bopt \sigmat \bdt - \bdt \bopt \sigmat - \phi^{*}(2 t)  ( \bopt \sigmat \bopt - \bopt^{2} \sigmat ) \right) \left( \frac{1}{2}\Gamma'_{-} + \frac{1}{2}\Gamma_{-}  + i\Delta_{-} + i\Delta'_{-} \right) \nonumber \\
+ &  \left(\bopt \sigmat \bdt - \sigmat \bdt \bopt - \phi(2 t)   ( \bdt \sigmat \bdt - \sigmat \bdtp{2}  ) \right) \left( \frac{1}{2}\Gamma'_{+} - i\Delta'_{+} \right) \nonumber \\
+ &  \left( \bdt \sigmat \bopt - \sigmat \bopt \bdt  - \phi^{*}(2 t)  ( \bopt \sigmat \bopt - \sigmat \bopt^{2} ) \right) \left( \frac{1}{2}\Gamma'_{-} + i\Delta'_{-} \right) + h.c,
\label{eq:MEgen}
\end{align}
We are still unable to see the physical relevance of the constants introduced from the above master equation. We can further simplify this equation by choosing the field to be in the vacuum state. This is also the case most relevant to our study. We consider this by setting $\langle n_{\kb} \rangle = 0$. This in turn results in $\Gamma'_{-} = \Gamma'_{+} = \Delta'_{-} = \Delta'_{+} = 0$. We note that all terms that remain finite after such a simplification on Eq. \eref{eq:MEgen}, are proportional to $\langle[\aop, \ad ] \rangle =1$. 
\begin{align}
\frac{\rmd \widetilde{\sigma}}{\rmd t} =  & \left( \bdt \sigmat \bopt - \bopt \bdt \sigmat - \phi(2 t)  ( \bdt \sigmat \bdt - \bdtp{2} \sigmat ) \right) \left( \frac{1}{2}\Gamma_{+}  - i\Delta_{+} \right)   \nonumber \\
+ &  \left( \bopt \sigmat \bdt - \bdt \bopt \sigmat - \phi^{*}(2 t)  ( \bopt \sigmat \bopt - \bopt^{2} \sigmat ) \right) \left( \frac{1}{2}\Gamma_{-}  + i\Delta_{-} \right) \nonumber\\
+ &\left( \bdt \sigmat \bopt - \sigmat \bopt \bdt - \phi^{\ast}(2 t)  ( \bopt \sigmat \bopt - \sigmat \bopt^{2} ) \right) \left( \frac{1}{2}\Gamma_{+}  + i\Delta_{+} \right)   \nonumber \\
+ &  \left( \bopt \sigmat \bdt - \sigmat \bdt \bopt - \phi(2 t)  ( \bdt \sigmat \bdt - \sigmat \bdtp{2} ) \right) \left( \frac{1}{2}\Gamma_{-}  - i\Delta_{-} \right),
\label{eq:MEprefinal}
\end{align}
This is clearly in the form of the general Redfield equations. So the master equation cannot be completely reduced to only contributions from Lindbladian terms.
However, rewriting it as 
\begin{align}
\frac{\rmd \widetilde{\sigma}}{\rmd t} =  & \Gamma_{+} \left( \bdt \sigmat \bopt - \frac{1}{2} \bopt \bdt \sigmat - \frac{1}{2} \sigmat \bopt \bdt  \right)  + \Gamma_{-}\left( \bopt \sigmat \bdt - \frac{1}{2} \bdt \bopt \sigmat - \frac{1}{2} \sigmat \bdt \bopt \right) \nonumber \\
-&\frac{1}{2}\Gamma_{+} \left(  \phi^{\ast}(2 t)  ( \bopt \sigmat \bopt - \sigmat \bopt^{2} ) + \phi(2 t)  ( \bdt \sigmat \bdt - \bdtp{2} \sigmat ) \right)  \nonumber \\
- & \frac{1}{2}\Gamma_{-} \left(  \phi(2 t)  ( \bdt \sigmat \bdt - \sigmat \bdtp{2} ) + \phi^{*}(2 t)  ( \bopt \sigmat \bopt - \bopt^{2} \sigmat ) \right) \nonumber\\
+ & i\Delta_{+} \left(\bopt \bdt \sigmat - \sigmat \bopt \bdt - \phi^{\ast}(2 t)  ( \bopt \sigmat \bopt - \sigmat \bopt^{2} ) + \phi(2 t)  ( \bdt \sigmat \bdt - \bdtp{2} \sigmat ) \right)  \nonumber \\
- &  i\Delta_{-}  \left(\bdt \bopt \sigmat - \sigmat \bdt \bopt - \phi(2 t)  ( \bdt \sigmat \bdt - \sigmat \bdtp{2} ) + \phi^{*}(2 t)  ( \bopt \sigmat \bopt - \bopt^{2} \sigmat ) \right),
\end{align}
we see that there are some terms that take the form of Lindbladian with $\Gamma_{+}$ and $\Gamma_{-}$ as the strengths of gain and damping respectively. Also from the first two terms of the contributions from $\Delta_{+}$ and $\Delta_{-}$, we can see that they will cause detuning when we return to the Schr{\"o}dinger picture. In order to evaluate the effective rates of damping, gain and detuning in the system, we need to evaluate the summation over $\mathbf{k}$ in (\ref{eq:constants1}) - (\ref{eq:constants4}). In a typical setup of our interest, the interaction with the reservoir is mainly due to vacuum fluctuations in an infinite range of field modes. So, it is more relevant to consider an integral over $\mathbf{k}$ in place of the summation, as indicated below. 
\begin{equation}
\label{eq:ksum2}
\sum_{\mathbf{k}} f(\mathbf{k}) \rightarrow \frac{V}{8 \pi^{3}}\int_{\mathbf{k}} f(\mathbf{k}) d^{3} \mathbf{k}.
\end{equation}
Using this along with Eq. (\ref{eq:gkApp}) in Eq. (\ref{eqn:constset}) and defining $(\mathbf{s}_{\mathbf{k}, \mu}\cdot \mathbf{e}_{x})$ where $\theta_{\kb,\mu}$ is the angle between $\mathbf{s}_{\mathbf{k}, \mu}$ and $\mathbf{e}_{x}$, we get, for instance, 
\begin{align}
    \Gamma_{\pm}&=\frac{1}{4 \pi^2 \epsilon_{0}} \, \frac{q^{2} \omega_{c} }{m c} \int_{\mathbf{k}}  \text{d}^{3} \mathbf{k} \, \frac{\delta(\omega_{c} \pm c k)}{k} \sin{(\theta_{\mathbf{k},\mu})} \cos^{2}{(\theta_{\mathbf{k},\mu})} \nonumber \\
    &=\frac{1}{4 \pi^2 \epsilon_{0}} \, \frac{q^{2} \omega_{c} }{m c^{2}} \int_{0}^{k_{\text{max}}} \text{d}k \, k^{2} \, \frac{\delta(k \pm \omega_{c}/c)}{k} \int_{0}^{\pi} \text{d}\theta_{\kb,\mu} \,  \sin{(\theta_{\kb,\mu})} \, \cos^{2}{(\theta_{\kb,\mu})} \int_{0}^{2 \pi} \text{d}\varphi \nonumber \\
    \implies \Gamma_{+}&=0 \quad \text{and}\\
    \Gamma_{-}&=\frac{1}{4 \pi \epsilon_{0}} \, \frac{4 \, q^{2} \omega_{c}^{2} }{3 \, m c^{3}} \quad (\text{provided } \omega_{c}/c \leqslant k_{\text{max}}). 
    \label{eq:GammaPM}
\end{align} 
Similarly,
\begin{align}
    \Delta_{\pm}&=\frac{1}{8 \pi^3 \epsilon_{0}} \, \frac{q^{2} \omega_{c} }{m c} \int_{\mathbf{k}}  \text{d}^{3} \mathbf{k} \, \frac{1}{k \, (\omega_{c} \pm c k)} \sin{(\theta_{\mathbf{k},\mu})} \cos^{2}{(\theta_{\mathbf{k},\mu})} \nonumber \\
    &=\frac{1}{8 \pi^3 \epsilon_{0}} \, \frac{q^{2} \omega_{c} }{m c^{2}} \int_{0}^{k_{\text{max}}} \text{d}k \, k^{2} \, \frac{c}{k \, (\omega_{c} \pm c k)} \int_{0}^{\pi} \text{d}\theta_{\kb,\mu} \,  \sin(\theta_{\kb,\mu}) \, \cos^{2}{(\theta_{\kb,\mu})} \int_{0}^{2 \pi} \text{d}\varphi \nonumber \\
    \implies \Delta_{\pm} &= \frac{\Gamma_{-}}{2 \pi \omega_{c}} \left(\pm \wmax - \omega_{c} \ln \left|\frac{\omega_{c} \pm \wmax}{\omega_{c}} \right| \right)
    \label{eq:deltapm2}
\end{align}
where $\wmax = c \kmax$.
As $\Gamma_{+} = 0$, the subscript is dropped from $\Gamma_{-}$ such that $\Gamma_{-} = \Gamma$. 
Using Eq. \eref{eq:MEprefinal}, the master equation now takes the form
\begin{align}
\frac{\text{d} \widetilde{\sigma}}{\text{d} t} = &   i \Delta_{+} \left( \bopt \bdt \sigmat - \bdt \sigmat \bopt + \phi(2t) ( \bdt \sigmat \bdt - \bdtp{2} \sigmat ) \right) \nonumber \\
+ &  \left( \bopt \sigmat \bdt - \bdt \bopt \sigmat + \phi^{*}(2t)  (\bopt^{2} \sigmat - \bopt \sigmat \bopt ) \right) \left( \frac{1}{2}\Gamma + i \Delta_{-} \right)  \nonumber \\
- & i \Delta_{+} \left( \sigmat \bdt \bopt - \bdt \sigmat \bopt + \phi^{*}(2t) ( \bopt \sigmat \bopt - \sigmat \bopt^{2} ) \right) \nonumber \\
+ &  \left( \bopt \sigmat \bdt - \sigmat \bopt \bdt + \phi(2t)  ( \sigmat \bdtp{2} - \bdt \sigmat \bdt ) \right) \left( \frac{1}{2}\Gamma - i \Delta_{-} \right)
\end{align}
Returning to the Schr{\"o}dinger picture and rearranging terms, the master equation becomes (Eq. \eref{eq:mastereq})
\begin{align}
    \frac{\text{d} \hat{\sigma}}{\text{d}t} = - &  i \left( \omega_{c} + \Delta_{-} - \Delta_{+} \right) \left[ \bd \bop , \hat{\sigma} \right] \nonumber \\
    + & \Gamma \left( \bop \hat{\sigma} \bd - \frac{1}{2}\bd \bop \hat{\sigma} - \frac{1}{2}\hat{\sigma} \bd \bop \right) \nonumber \\ 
    - & i \Delta_{+} \left( \bop \hat{\sigma} \bop - \hat{\sigma} \bop^{2} \right) \nonumber - \left( \frac{1}{2}\Gamma + i \Delta_{-} \right) \left( \bop \hat{\sigma} \bop - \bop^{2} \hat{\sigma} \right)  \nonumber \\ 
    + &  i \Delta_{+} \left( \bd \hat{\sigma} \bd -  \bdp{2} \hat{\sigma} \right) - \left( \frac{1}{2}\Gamma -  i \Delta_{-} \right) \left( \bd \hat{\sigma} \bd - \hat{\sigma} \bdp{2} \right).
\end{align}
This master equation in terms of $\hat{x}$ and $\hat{p}_{x}$ is
\begin{align}
    \frac{\text{d} \hat{\sigma}}{\text{d}t} = - &  \frac{i}{\hbar} \left( 1 + \frac{2\left(\Delta_{-} - \Delta_{+}\right)}{\omega_{c}} \right) \left[\frac{\hat{p}^{2}_{x}}{2m}, \hat{\sigma} \right]
    - \frac{i \omega_{c}^{2} m}{2 \hbar} \left[ \hat{x}^{2}, \hat{\sigma} \right] \nonumber \\ 
    + & \frac{\Gamma}{ \hbar m \omega_{c}} \left( \hat{p}_{x} \hat{\sigma} \hat{p}_{x} -\frac{1}{2} \hat{p}_{x}^{2} \hat{\sigma} - \frac{1}{2} \hat{\sigma} \hat{p}_{x}^{2}  \right) \nonumber \\
    + & \frac{1}{\hbar}\left( \Delta_{-} + \Delta_{+} + \frac{i\Gamma}{2} \right) \left(\hat{p}_{x} \hat{\sigma} \hat{x} - \frac{1}{2} \hat{x} \hat{p}_{x} \hat{\sigma} - \frac{1}{2} \hat{\sigma} \hat{x} \hat{p}_{x} \right)  \nonumber \\
    + & \frac{1}{\hbar}\left( \Delta_{-} + \Delta_{+} - \frac{i\Gamma}{2} \right)\left(\hat{x} \hat{\sigma} \hat{p}_{x} - \frac{1}{2} \hat{p}_{x} \hat{x} \hat{\sigma} - \frac{1}{2} \hat{\sigma} \hat{p}_{x} \hat{x} \right) \nonumber \\
    + & \frac{1}{2\hbar}\left(\Delta_{-} - \Delta_{+} + \omega_{c} - \frac{i \Gamma}{2} \right) \left[\hat{\sigma}, \hat{p}_{x} \hat{x} \right] \nonumber \\
    + & \frac{1}{2\hbar}\left(\Delta_{-} - \Delta_{+} + \omega_{c} + \frac{i \Gamma}{2} \right) \left[\hat{x} \hat{p}_{x} , \hat{\sigma} \right].
\end{align}

\section{\label{app:FreePart} Free Particle}

In this section, we study the dynamics of a free particle with an electric charge $q$ and mass $m$ in the presence of vacuum fluctuations of the EM field. This is achieved by following a procedure similar to that used in obtaining the master equation for a trapped charged particle as outlined in Supplementary Material Sec. \ref{app:MEderiv}. The interaction with the electromagnetic field is via the term proportional to $\mathbf{p}.\mathbf{A}$ where $\mathbf{p}$ is the momentum of the particle and $\mathbf{A}(\mathbf{r},t)$ is the magnetic vector potential. The effective Hamiltonian for the free particle interacting with the EM field is 
\begin{align}
\label{eq:HFP1}
 H_{0} = &\frac{\left(\mathbf{p}- q\mathbf{A}(\mathbf{r},t)\right)^{2}}{2m}
    + \frac{\epsilon_{0}}{2} \iiint_{V} \rmd^{3}\mathbf{r} \left( |\mathbf{E}(\mathbf{r},t)|^{2} + c^{2}  |\mathbf{B}(\mathbf{r},t)|^{2}\right).
\end{align}
We note that this differs from the Hamiltonian $\hat{H}_{\textsc{p}}$ in the main body of the paper by the absence of the harmonic potential term.
As pointed out in the letter, expanding the square in the first term of the Hamiltonian results in two interaction terms, one dependent on $\mathbf{p} \cdot \mathbf{A}(\mathbf{r},t)$ and the other on $\mathbf{A}^{2}(\mathbf{r},t)$. Here too we make the assumption that the term dependent on $\mathbf{A}^{2}(\mathbf{r},t)$ is much smaller in magnitude and is therefore neglected, to give
\begin{align}
\label{eq:HFP2}
 H_{0} = &\frac{\mathbf{p}^{2}}{2m} - \frac{q}{m} \mathbf{p} \cdot \mathbf{A}(\mathbf{r},t)
    + \frac{\epsilon_{0}}{2} \iiint_{V} \rmd^{3}\mathbf{r} \left( |\mathbf{E}(\mathbf{r},t)|^{2} + c^{2}  |\mathbf{B}(\mathbf{r},t)|^{2}\right).
\end{align}
In the semi-classical limit, the free particle Hamiltonian is the standard
\begin{align}
    \hat{H}_{\textsc{p} 0} = \frac{\hat{p}^{2}}{2m}.
\end{align}
As in the earlier case, we move to a fully quantum approach by considering a quantum EM field using,
\begin{equation}
    \label{eq:afieldb}
    \mathbf{\hat{A}}(\mathbf{r},t) = \sum_{\mathbf{k}, \mu} \sqrt{\frac{\hbar}{2 \Omega_{\mathbf{k}} \epsilon_{0} V}} \mathbf{s}_{\mathbf{k}, \mu} \left(\hat{a}_{\mathbf{k}}(t) e^{i \mathbf{k} \cdot \mathbf{r}} + \hat{a}^{\dagger}_{\mathbf{k}}(t) e^{-i \mathbf{k} \cdot \mathbf{r}}  \right).
\end{equation}
We recall $\aop$ and $\ad$ are the photon annihilation and creation operators, respectively, for each mode of the field, characterised by the wave vector $\kb$ and polarisation vector $\mathbf{s}_{\mathbf{k}, \mu}$. We also recall that the mode frequency is $\Omega_{\kb} = c \kb $ and $V$ is any arbitrary normalisation volume.

Substituting Eq. \eref{eq:afieldb} into the Hamiltonian \eref{eq:HFP2}, the fully quantum Hamiltonian becomes
\begin{align}
    \hat{H}_{0} = \hat{H}_{\textsc{p} 0} + \hat{H}_{\textsc{i} 0} + \hat{H}_{\textsc{f} 0}
\end{align}
where
\begin{subequations}
\begin{align}
    \label{eq:HF0}
    &\hat{H}_{\textsc{f} 0} = \sum_{\kb,\mu} \hbar \Omega_{\kb} \left( \ad \aop + \frac{1}{2} \right),\\
    \label{eq:HI0}
    &\hat{H}_{\textsc{i} 0} =  \sum_{\kb,\mu} f_{\kb,\mu} \hat{p} \left( \aop  + \ad \right),
\end{align}
\label{eq:Hfi0}
\end{subequations}
and
\begin{align}
    f_{\kb,\mu} &= - \left( \frac{q}{m} \sqrt{\frac{\hbar }{2 \epsilon_{0} V \wk}} \right) \sin{(\theta_{\kb, \mu})} \cos{(\phi_{\mathbf{k}, \mu})}.
\end{align}
We recall that $\theta_{\kb,\mu}$ and $\phi_{\mathbf{k}, \mu}$ are the standard polar angles. As discussed in Appendix \ref{app:MEderiv}, we account for the contributions from either polarisations by
\begin{align}
    f_{\kb} &= 2 f_{\kb, \mu} = - \left( \frac{q}{m} \sqrt{\frac{2 \hbar }{ \epsilon_{0} V \wk}} \right) \sin{(\theta_{\kb})} \cos{(\phi_{\mathbf{k}})}.
\end{align}

As before, we want to study the dynamics of the particle subsystem using its time-evolved reduced density matrix $\hat{\sigma}(t) = \text{Tr}_{\textsc{f}}[ \rho(t) ]$ where $\rho(t)$ denotes the density matrix of the full system. The density matrix corresponding to the field $\hat{\sigma}_{\textsc{f}} (t) = \text{Tr}_{\textsc{p}}[ \rho(t) ]$. As pointed out in Appendix \ref{app:MEderiv}, the stationarity of the field subsystem leads to $\hat{\sigma}_{\textsc{f}}$ being independent of $t$. 
Moving to the interaction picture, we find that
\begin{subequations}
\begin{align}
    \label{eq:pint}
    \tilde{p} (t) &= \hat{p} (0) \equiv \hat{p} \\
    \label{eq:afpint}
    \aopt (t) &= \aop(0) e^{- i \Omega_{\kb} t} \equiv \aop e^{-i \wk t}.
\end{align}
\label{eq:fpint}
\end{subequations}
Following the same approach outlined in Appendix \ref{app:MEderiv}, we find the master equation for a free particle to be
\begin{align}
\label{eq:mastereqfp}
\frac{d\hat{\sigma}_{\textsc{fp}}}{dt} = \frac{i}{\hbar} \left(1 - \delta E_{\textsc{fp}} \right) \left[\hat{\sigma}, \hat{H}_{\textsc{p}0} \right],
\end{align}
where
\begin{align}
    \delta E_{\textsc{fp}}&=  \frac{2 q^{2}}{3 \pi^2 \epsilon_{0} m c^{3}} \; \wmax
    \label{eq:deltafp}
\end{align}

From (\ref{eq:mastereqfp}), it can be seen that $\delta E_{\textsc{fp}}$ corresponds to a global shift in the energy of the free particle which is not observable. This shift can be interpreted as a renormalisation of the mass of the particle due to the presence of the vacuum state of the field. This global shift is present both in the case of a free particle and a trapped particle. We need to regularize the results obtained by adding appropriate counter-terms, both in the case of the free and the trapped particle~\cite{Bethe}.

In order to regularise the results in the case of the harmonically trapped particle, the master equation (\ref{eq:mastereq}) in that case is recalled here and is also rewritten as,
\begin{align}
    \label{eq:masterfirstline}
    \frac{d\hat{\sigma}}{dt} &= \frac{i}{\hbar}\left(1 - \frac{\Delta_{+} - \Delta_{-}}{\omega_{c}}\right) \left[ \hat{\sigma}, 
    H_{P} \right] \nonumber \\
    + & \Gamma \left( \bop \hat{\sigma} \bd - \frac{1}{2}\bd \bop \hat{\sigma} - \frac{1}{2}\hat{\sigma} \bd \bop \right) \nonumber \\ 
    - & i \Delta_{+} \left( \bop \hat{\sigma} \bop - \hat{\sigma} \bop^{2} \right) \nonumber - \left( \frac{1}{2}\Gamma + i \Delta_{-} \right) \left( \bop \hat{\sigma} \bop - \bop^{2} \hat{\sigma} \right)  \nonumber \\ 
    + &  i \Delta_{+} \left( \bd \hat{\sigma} \bd -  \bdp{2} \hat{\sigma} \right) - \left( \frac{1}{2}\Gamma -  i \Delta_{-} \right) \left( \bd \hat{\sigma} \bd - \hat{\sigma} \bdp{2} \right).
\end{align}
The energy shift in the case of the harmonic oscillator is 
\begin{align}
    \delta E = \frac{\Delta_{+} - \Delta_{-}}{\omega_{c}}.
\end{align}
Using the definition of the parameters $\Delta_{\pm}$ (Eq. \eref{eq:deltapm2}), we see that the shift has two terms: one linear in $\wmax$ and one logarithmic in $\wmax$. Considering the linear term $\delta E_{\text{lin}}$ in the shift $\delta E$, 
\begin{align}
    \delta E_{\text{lin}} = \frac{q^{2}}{3 \pi^{2} \epsilon_{0} m c^{3}} \wmax.
\end{align}
It is interesting to note that 
\begin{align}
   \frac{\delta E_{\textsc{fp}}}{2} =  \delta E_{\text{lin}}.
\end{align}
This can be explained as follows.
Bethe argues that the regularisation of the energy shift in the trapped case is achieved by subtracting the energy shift of a free particle, choosing the kinetic energy of that free particle to be the same as the average kinetic energy of the trapped system~\cite{Bethe}. As our particle is in a harmonic trap, its average kinetic energy is just half the total energy of the system. We, therefore, see that the $\delta E_{\text{lin}}$ is the term that needs to be subtracted out to regularise the results. 
The shift $\delta E$ in the case of the harmonic oscillator is left solely with the logarithmic dependence in $\wmax$. Alternatively, this can also be achieved by introducing the appropriate counter-terms to the Hamiltonian as described in \cite{Welton}.

\section{\label{app:validity} Validity of Master Equation}

The two conditions that must be satisfied to ensure the validity of the time-evolved state $\hat{\sigma}(t)$ (obtained by solving the master equation \eref{eq:mastereq2} with the renormalized constants) are as follows.
\begin{enumerate}
    \item The trace of $\hat{\sigma}$ is preserved.
    \item $\hat{\sigma}$ is a completely positive matrix.
\end{enumerate}
The trace of $\hat{\sigma}$ is easily checked by taking the trace of both sides of the master equation (\ref{eq:mastereq2})
\begin{align}
    \text{Tr}\left[\frac{\text{d} \hat{\sigma}}{\text{d}t} \right] =  - &  i \left( \omega_{c} + \Delta_{-}^{\textsc{(r)}} - \Delta_{+}^{\textsc{(r)}} \right) \left( \text{Tr} \left[ \bd \bop \hat{\sigma} \right] - \text{Tr} \left[ \hat{\sigma} \bd \bop \right] \right)  \nonumber \\
    + & \Gamma \left( \text{Tr} \left[ \bop \hat{\sigma} \bd \right] - \frac{1}{2} \text{Tr} \left[  \bd \bop \hat{\sigma} \right] - \frac{1}{2} \text{Tr} \left[ \hat{\sigma} \bd \bop \right] \right), \nonumber \\ 
    - & \Delta_{+}^{\textsc{(r)}} \left(\text{Tr} \left[ \bop \hat{\sigma} \bop \right] + \text{Tr} \left[ \hat{\sigma} \bop^{2} \right]  \right) - \left( \frac{1}{2}\Gamma + i \Delta_{-}^{\textsc{(r)}} \right) \left( \text{Tr} \left[ \bop \hat{\sigma} \bop \right] - \left[ \bop^{2} \hat{\sigma} \right] \right)  \nonumber \\ 
    + &  i \Delta_{+}^{\textsc{(r)}} \left( \left[ \bd \hat{\sigma} \bd \right] -  \left[ \bdp{2} \hat{\sigma} \right] \right) - \left( \frac{1}{2}\Gamma -  i \Delta_{-}^{\textsc{(r)}} \right) \left( \left[ \bd \hat{\sigma} \bd \right] - \left[ \hat{\sigma} \bdp{2} \right] \right).
\end{align}
Using the invariance of cyclic permutations of the trace it can be seen
\begin{align}
    \text{Tr}\left[\frac{\text{d} \hat{\sigma}}{\text{d}t} \right] = 0,
\end{align}
and hence the trace is preserved.

The second condition is guaranteed to be satisfied if the Husimi Q-function corresponding to the master equation remains a Gaussian at every instant of time if it is initially in a Gaussian state~\cite{Adesso2014}. Here the Q function corresponding to the state $\hat{\sigma}(t)$ is defined as
\begin{align}
    \label{eq:qfunction}
    Q (\alpha, \alpha^{*}, t ) = \frac{1}{\pi} \bra{\alpha} \hat{\sigma} (t) \ket{\alpha},
\end{align}
where $\ket{\alpha}$ is the standard coherent state of the particle system, given by
\begin{equation}
    \ket{\alpha} = e^{-|\alpha|^{2}/2} \sum_{p=0}^{\infty} \frac{\alpha^{p}}{p!} b^{\dagger p} \ket{0}
\end{equation}
with $\ket{0}$ being the ground state of the particle in the oscillator potential.

By substituting Eq. \eref{eq:qfunction} into the master equation \eref{eq:mastereq}, the following differential equation for $Q (\alpha, \alpha^{*}, t )$ is obtained.
\begin{align}
    \label{eq:qfunction2}
   \frac{\text{d}}{\text{d}t}Q(\alpha, \alpha^*, t) &= \frac{\Gamma}{2} \frac{\partial^{2}Q}{\partial \alpha \partial \alpha^{*}} + A_{1} \frac{\partial^{2}Q}{\partial \alpha^{2}} + A_{2} \frac{\partial Q}{\partial \alpha} + \frac{\Gamma}{2} Q + c.c,
\end{align}
where c.c is the complex conjugate and
\begin{align}
    A_{1} &= \left(\frac{\Gamma}{2} - i \Delta_{-}^{\textsc{(r)}} \right), \\
    A_{2} &= i \alpha \omega_{c} + \left( \alpha - \alpha^{*} \right) \left( \Delta_{-}^{\textsc{(r)}} - \Delta_{+}^{\textsc{(r)}} \right) +\frac{\Gamma}{2}\left( \alpha + \alpha^{*} \right).
\end{align}
The Q function is related to the characteristic function $\widetilde{W} (\xi, \xi^{*}, t)$ through~\cite{agarwal2012}
\begin{align}
    \label{eq:charac}
    Q (\alpha, \alpha^{*}, t ) &= \frac{1}{\pi^{2}} \int \widetilde{W} (\xi, \xi^{*}, t) e^{\alpha \xi^{*} - \alpha^{*} \xi} \; \text{d}^{2}\xi.
\end{align}
Substituting Eq. \eref{eq:charac} into Eq. \eref{eq:qfunction2} gives
\begin{align}
     \label{eq:charac2}
    \int \; \text{d}^{2}\xi' \; \frac{\text{d}}{\text{d}t}\widetilde{W} (\xi', \xi'^{*}, t) &= \int \; \text{d}^{2}\xi \; \Big[ i \omega_{c} \left( \alpha^{*} \xi + \alpha \xi^{*} \right) + i \left( \Delta_{-}^{\textsc{(r)}} - \Delta_{+}^{\textsc{(r)}} \right) \left( \alpha^{*} \xi + \alpha \xi^{*} - \alpha^{*} \xi^{*} - \alpha \xi \right) \nonumber \\
    &+ \frac{\Gamma}{2} \left( \alpha^{*} \xi^{*} + \alpha \xi^{*}- \alpha \xi - \alpha^{*} \xi + \xi^{2} + \xi^{* 2} - |\xi|^{2} + 2 \right) + i \Delta_{-}^{\textsc{(r)}} \left(\xi^{2} - \xi^{* 2} \right) \Big] \widetilde{W} (\xi, \xi^{*}, t).  
\end{align}
Equation \eref{eq:charac2} must remain true for any choice of $\xi$, therefore equating the integrands gives
\begin{align}
    \label{eq:charac3}
    \frac{\text{d}}{\text{d}t}\widetilde{W} (\xi, \xi^{*}, t) &= w(\alpha, \alpha^{*}, \xi, \xi^{*}) \widetilde{W} (\xi, \xi^{*}, t),  
\end{align}
where
\begin{align}
    w(\alpha, \alpha^{*}, \xi, \xi^{*})&= \Big[ i \omega_{c} \left( \alpha^{*} \xi + \alpha \xi^{*} \right) + i \left( \Delta_{-}^{\textsc{(r)}} - \Delta_{+}^{\textsc{(r)}} \right) \left( \alpha^{*} \xi + \alpha \xi^{*} - (\alpha^{*} \xi^{*} + \alpha \xi) \right) \nonumber \\
    &+ \frac{\Gamma}{2} \left( \alpha^{*} \xi^{*} + \alpha \xi^{*}- \alpha \xi - \alpha^{*} \xi + \xi^{2} + \xi^{* 2} - 2 |\xi|^{2} + 2 \right) + i \Delta_{-}^{\textsc{(r)}} \left(\xi^{2} - \xi^{* 2} \right) \Big].
 \end{align}
From Eq. \eref{eq:charac3}, we find
\begin{align}
    \label{eq:charac4}
    \widetilde{W} (\xi, \xi^{*}, t) &= \widetilde{W}_{0} e^{w t},
\end{align}
where $\widetilde{W}_{0}$ is the initial characteristic function. We are interested in determining the evolution for initial coherent states $\ket{\alpha_{0}}$. The initial characteristic function corresponding to the standard coherent state is
\begin{align}
    \label{eq:winit}
    \widetilde{W}^{\alpha}_{0} = e^{-|\xi|^{2} + \xi \alpha_{0}^{*} - \xi^{*} \alpha_{0}}.
\end{align}
Substituting Eqs. \eref{eq:charac4} and \eref{eq:winit} into Eq. \eref{eq:charac}, we obtain
\begin{align}
     Q (\alpha, \alpha^{*}, t ) &= \frac{1}{\pi^{2}} \int \; \text{d}^{2} \xi \;  e^{-|\xi|^{2} + \xi \alpha_{0}^{*} - \xi^{*} \alpha_{0} + \alpha \xi^{*} - \alpha^{*} \xi} e^{w t}.
\end{align}
Defining the real and imaginary parts of $\alpha$ and $\xi$,
\begin{align}
    \alpha &= \alpha_{\textsc{Re}} + i \alpha_{\textsc{Im}}, \\
    \xi &= \xi_{\textsc{Re}} + i \xi_{\textsc{Im}},
\end{align}
the Q-function is
\begin{align}
    \label{eq:qint2}
     Q (\alpha, \alpha^{*}, t ) &= \frac{1}{\pi^{2}} \int \; \text{d} \xi_{\textsc{Re}} \text{d} \xi_{\textsc{Im}} \;  e^{\Upsilon (\xi, \xi^{*})}.
\end{align}
Here
\begin{align}
    \Upsilon (\xi, \xi^{*}) &= \left[ 2 i \omega_{c} \left( \ar \zr + \ai \zi \right) + 4 i \left( \Delta_{-}^{\textsc{(r)}} - \Delta_{+}^{\textsc{(r)}} \right) \ai \zi + \Gamma \left(1 - 2 \zi^{2} - 2 i \ar \zi \right) - 4 \Delta_{-}^{\textsc{(r)}} \zr \zi \right] t \nonumber \\
    &- \zi^{2} - \zr^{2} + 2 i \left( \zi \aor - \zr \aoi \right) + 2 i \left( \ai \zr - \ar \zi \right).
\end{align}
Considering only the integral corresponding to the variable $\zi$ in Eq. \eref{eq:qint2},
\begin{align}
    \nonumber \int\limits_{-\infty}^{\infty} \rmd \zi \, \exp \left[ - (1+2 \Gamma t)  \zi^{2} + 2 i \left( \zi \aor - \ar \zi + \omega_{c} t \ai \zi + 2 (\Delta \omega_{c}) t \ai \zi - \Gamma t \ar \zi + 2 i \Delta_{-}^{\textsc{(r)}} t \zr \zi\right)\right]\\
    =\sqrt{\frac{\pi}{1+2 \Gamma t}} \exp \left[ -\frac{1}{1+2\Gamma t} \left( \aor - \ar+\omega_{c} t \ai + 2 (\Delta \omega_{c}) t \ai - \Gamma t \ar +2 i \Delta_{-}^{\textsc{(r)}} t \zr \right)^{2}\right] \label{eq:integ1}
\end{align}
where we recall $\Delta \omega_{c}=\Delta_{-}^{\textsc{(r)}}-\Delta_{+}^{\textsc{(r)}}$.
Using Eq. \eref{eq:integ1} in Eq. \eref{eq:qint2},
\begin{align}
Q(\ar,\ai,t)=\frac{e^{\Gamma t}}{\pi\sqrt{\pi (1+2 \Gamma t)}} \exp \left[- \frac{1}{1+2\Gamma t} \left( \aor - \ar+\omega_{c} t \ai + 2 (\Delta \omega_{c}) t \ai - \Gamma t \ar \right)^{2}\right] \nonumber\\
\int\limits_{-\infty}^{\infty} \rmd \zr \, \exp\left[- \left(1- \frac{4 \left(\Delta_{-}^{\textsc{(r)}}\right)^{2} t^{2}}{1+2\Gamma t}\right)\zr^{2}+ 2 i \left( \ai \zr  - \aoi \zr  + \omega_{c} t \ar \zr \right) \right]\nonumber \\
 \exp \left[ -\frac{4 i \Delta_{-}^{\textsc{(r)}} t \zr}{1+2\Gamma t} \left( \aor - \ar+\omega_{c} t \ai + 2 (\Delta \omega_{c}) t \ai - \Gamma t \ar \right)\right].
 \label{eq:QFuncFinal}
\end{align}
This integral in Eq. \eref{eq:QFuncFinal} will also yield a positive Gaussian function, like the first in Eq. \eref{eq:integ1}, if and only if
\begin{align}
    \left(1- \frac{4 \left(\Delta_{-}^{\textsc{(r)}}\right)^{2} t^{2}}{1+2\Gamma t}\right) > 0,\\
    \text{equivalently, } \Gamma - 2 \left(\Delta_{-}^{\textsc{(r)}}\right)^{2} t > -\frac{1}{2 t}.\label{eq:condnalph}
\end{align}
Therefore, our master equation  \eref{eq:mastereq} corresponds to a valid mapping of states from $t=0$ to $t=T_{\text{max}}$ where
\begin{align}
    T_{\text{max}} = \frac{\Gamma + \sqrt{\Gamma^{2} + 4 \left(\Delta_{-}^{\textsc{(r)}}\right)^{2}}}{4 \left(\Delta_{-}^{\textsc{(r)}}\right)^{2}}.
    \label{eq:tmax}
\end{align}
It can be seen for the typical experimental parameter values in Fig.  \ref{fig:modelsketch} in the letter, $T_{\text{max}} = 0.04s$, where the cut-off $\Omega_{3}$ has been chosen as this gives the lowest possible $T_{\text{max}}$. 

We have shown that the Q - function remains positive for initial coherent states for a limited time $T_{\text{max}}$. We now wish to do the same for initial thermal states. The thermal state density operator is defined as
\begin{align}
    \hat{\sigma}_{\textsc{th}} = \frac{1}{Z} \sum_{n} e^{- n \beta \hbar \omega_{c}} \ket{n} \bra{n},
\end{align}
where $\beta$ is the inverse temperature $1/k_{B} T$ and $k_{B}$ is the Boltzmann constant.

The Q - function corresponding to an initial thermal state is
\begin{align}
    Q_{\textsc{th}} = \frac{1}{\pi Z} \sum_{n} e^{-n \beta \hbar \omega_{c}} e^{- |\alpha|^{2}} \frac{|\alpha|^{2n}}{n!}.
\end{align}
Using the Taylor expansion of $e^{|\alpha|^{2} e^{-\beta \hbar \omega_{c}}}$, the Q function now takes the form
\begin{align}
    Q_{\textsc{th}} = \frac{1}{\pi Z} e^{-|\alpha|^{2} (1- e^{-\beta \hbar \omega_{c}})}.
\end{align}
Rewriting the Q function in terms of $\alpha'=\alpha \sqrt{1-e^{-\beta \hbar \omega_{c}}}$, it is evident that the Q-function of the thermal state takes a form identical to that of the vacuum state \footnote{We repeat the steps listed above for the coherent state setting $\alpha_{0}$ to zero to ensure positivity of the Q function in the case of the vacuum state}.
Therefore an initial thermal state will also remain positive over the range $t=0$ to $t=T_{\text{max}}$.

\section{\label{app:expectxandp} Expectation value of Position}

In this section, we study the dynamics of the trapped particle through the expectation value of its position. Through this exercise, we gain insight into the effects on the particle due to each term in the master equation \eref{eq:mastereq2}. 

The temporal dynamics of the expectation value of any operator $\hat{\mathcal{O}}$ can be found using
\begin{align}
    \label{eq:expect}
    \frac{\text{d}\langle\hat{\mathcal{O}} \rangle}{\text{d}t} = \text{Tr}\left[ \frac{\text{d} \hat{\sigma}}{\text{d}t} \hat{\mathcal{O}}  \right],
\end{align}
obtained from the master equation \eref{eq:mastereq}.

Using (\ref{eq:expect}), differential equations for the position $\langle \hat{x} \rangle$, and momentum $\langle \hat{p} \rangle$  of the particle can be found
\begin{align}
    \label{eq:xexpect}
    \frac{\text{d}\langle\hat{x}\rangle}{\text{d}t} &= - \Gamma \langle \hat{x} \rangle + \frac{\omega_{c} + 2 \Delta \omega_{c}}{m \omega_{c}} \langle \hat{p} \rangle,  \\
    \label{eq:pexpect}
    \frac{\text{d}\langle\hat{p}\rangle}{\text{d}t} &= - m \omega_{c}^{2} \langle \hat{x} \rangle.
\end{align}
Taking the time derivative of (\ref{eq:xexpect}) and substituting (\ref{eq:pexpect}) gives a second order differential equation in $\langle \hat{x} \rangle$
\begin{align}
    \label{eq:2ndx}
    \frac{\text{d}^{2}\langle\hat{x}\rangle}{\text{d}t^{2}} + \Gamma \frac{\text{d}\langle\hat{x}\rangle}{\text{d}t} + \omega_{c} \left( \omega_{c} + 2 \Delta \omega_{c} \right) \langle \hat{x} \rangle = 0.
\end{align}
A general solution for (\ref{eq:2ndx}) is of the form
\begin{align}
    \langle \hat{x} \rangle (t) = x_0 \left( e^{t \lambda_{+}} + e^{t \lambda_{-}} \right),
\end{align}
where $x_{0}$ is the initial expectation value of the particle position and
\begin{align}
    \label{eq:lambda}
    \lambda_{\pm} = -\frac{\Gamma}{2} \pm \frac{\sqrt{\Gamma^{2} - 4 \omega_{c}^{2} - 8 \omega_{c} \Delta \omega_{c}}}{2}.
\end{align}
This can be rewritten as 
\begin{align}
    \lambda_{\pm} = -\frac{\Gamma}{2} \pm i \omega_{c} \sqrt{ 1 + \frac{2 \Delta \omega_{c}}{\omega_{c}} - \frac{\Gamma^{2}}{4 \omega_{c}^{2}}}.
\end{align}
Assuming $\Delta \omega_{c}$, $\Gamma \ll \omega_{c}$, a Taylor series expansion of the square root gives
\begin{align}
    \lambda_{\pm} = -\frac{\Gamma}{2} \pm i \omega_{c} \left( 1 + \frac{ \Delta \omega_{c}}{\omega_{c}} - \frac{\Gamma^{2}}{8 \omega_{c}^{2}} + \dots \right).
\end{align}
Retaining only $\Delta \omega_{c}/\omega_{c}$, 
\begin{align}
     \lambda_{\pm} \approx -\frac{\Gamma}{2} \pm i \left( \omega_{c}  +  \Delta \omega_{c} \right).
\end{align}
Therefore, in this limit, the solution for $\langle \hat{x} \rangle$ is
\begin{align}
    \langle \hat{x} \rangle (t) = x_{0} e^{-\frac{\Gamma}{2}t} \cos{\left[\left( \omega_{c} + \Delta \omega_{c} \right) t \right]}.
\end{align}
We see that this can be seen as a damped harmonic oscillator which has suffered a shift $\Delta \omega_{c}$ to its natural frequency. These effects at this limit are fully explained by merely using the first two lines of the following master equation (recalling Eq. \eref{eq:mastereq2}),
\begin{align*}
    \frac{\text{d} \hat{\sigma}}{\text{d}t} = - &  i \left( \omega_{c} + \Delta_{-}^{\textsc{(r)}} - \Delta_{+}^{\textsc{(r)}} \right) \left[ \bd \bop , \hat{\sigma} \right] \nonumber \\
    + & \Gamma \left( \bop \hat{\sigma} \bd - \frac{1}{2}\bd \bop \hat{\sigma} - \frac{1}{2}\hat{\sigma} \bd \bop \right) \nonumber \\ 
    - & i \Delta_{+}^{\textsc{(r)}} \left( \bop \hat{\sigma} \bop - \hat{\sigma} \bop^{2} \right) \nonumber - \left( \frac{1}{2}\Gamma + i \Delta_{-}^{\textsc{(r)}} \right) \left( \bop \hat{\sigma} \bop - \bop^{2} \hat{\sigma} \right)  \nonumber \\ 
    + &  i \Delta_{+}^{\textsc{(r)}} \left( \bd \hat{\sigma} \bd -  \bdp{2} \hat{\sigma} \right) - \left( \frac{1}{2}\Gamma -  i \Delta_{-}^{\textsc{(r)}} \right) \left( \bd \hat{\sigma} \bd - \hat{\sigma} \bdp{2} \right).
\end{align*}
The contributions from the other terms are noticibly smaller and negligible in the case of the first-order expectation value. Higher-order cumulants of the position observable might reveal more information on these other terms.

\section{\label{app:model} EM vacuum effects on the cyclotron motion}

We model the cyclotron motion, of frequency $\omega_{c}$, of an electron, of mass $m$ and charge $q$, confined to the $x$-$y$ plane of an SEC (see Fig. \ref{fig:modelsketch}) interacting with the vacuum modes of the EM field. The minimal coupling classical Hamiltonian, $H_{\text{tot}}$ in the Coulomb gauge is
\begin{align}
    H_{\text{tot}} =& H_{\text{cyc}} + H_{\textsc{f}},\\
    H_{\text{cyc}} = &\frac{|p_{x} \mathbf{e}_{x} + p_{y} \mathbf{e}_{y} - q\mathbf{A}(\mathbf{r},t)|^{2}}{2m} + \frac{m \omega_{c}^{2} x^{2}}{2}+ \frac{m \omega_{c}^{2} y^{2}}{2},\\
    H_{\textsc{f}}=& \frac{\epsilon_{0}}{2} \iiint_{V} \rmd^{3}\mathbf{r}' \left( |\mathbf{E}(\mathbf{r}',t)|^{2} + c^{2}  |\mathbf{B}(\mathbf{r}',t)|^{2}\right),
\end{align}
where $\mathbf{r}\equiv(x,y,z)$ denotes the position and $\mathbf{p}\equiv(p_{x},p_{y},p_{z})$ the corresponding canonical momentum of the particle. We note that the motion along the $z$-axis is constrained, setting $p_{z}=0$. Here $
\mathbf{e}_{i}$ ($i=x, y$) is the unit vector along the $i$-axis, $\mathbf{A}(\mathbf{r}, t)$ is the magnetic vector potential, 
and $\mathbf{E}(\mathbf{r},t)$ (respectively, $\mathbf{B}(\mathbf{r},t)$) is the electric (respectively, magnetic) field. We make use of the following assumption
\begin{enumerate}[label=(\roman*)]  
    \item \label{Ass:WeakField} \textit{Weak field limit}, $|\mathbf{A}(\mathbf{r}, t)|^{2} \ll \left\vert \dfrac{2}{q} \: \mathbf{p} \cdot \mathbf{A} (\mathbf{r}, t)\right\vert$,
\end{enumerate}
such that
\begin{align}
    H_{\text{cyc}} &= H_{x} + H_{y} \label{eq:cyclotron}\\ 
    H_{x}&=\frac{p_{x}^{2}}{2 m} - \frac{q}{m}p_{x}(\mathbf{A}(\mathbf{r}, t)\cdot\mathbf{e}_{x}) + \frac{m \omega_{c}^{2} x^{2}}{2} \\ 
    H_{y}&= \frac{p_{y}^{2}}{2 m} - \frac{q}{m}p_{y}(\mathbf{A}(\mathbf{r}, t)\cdot\mathbf{e}_{y}) + \frac{m \omega_{c}^{2}  y^{2}}{2}.    
\end{align}
We then quantise $H_{\text{tot}}$ where the $\mathbf{A}(\mathbf{r},t)$, $\mathbf{E}(\mathbf{r},t)$ and $\mathbf{B}(\mathbf{r},t)$ fields are as follows
\begin{subequations}
\begin{align}
    \label{eq:appAfield}
    \mathbf{\hat{A}}(\mathbf{r},t) &= \sum_{\mathbf{k}, \mu} \sqrt{\frac{\hbar}{2 \Omega_{\mathbf{k}} \epsilon_{0} V}} \mathbf{s}_{\mathbf{k}, \mu} \left(\hat{a}_{\mathbf{k}}(t) e^{i \mathbf{k} \cdot \mathbf{r}} + \hat{a}^{\dagger}_{\mathbf{k}}(t) e^{-i \mathbf{k} \cdot \mathbf{r}}  \right),\\
    \label{eq:appEfield}
    \mathbf{\hat{E}}(\mathbf{r},t) &= i \sum_{\mathbf{k}, \mu}  \mathbf{s}_{\mathbf{k}, \mu} \sqrt{\frac{\hbar \wk}{2 \epsilon_{0} V}} \left(\hat{a}_{\mathbf{k}}(t) e^{i \mathbf{k} \cdot \mathbf{r}} - \hat{a}^{\dagger}_{\mathbf{k}}(t) e^{-i \mathbf{k} \cdot \mathbf{r}}  \right) \\
    \label{eq:appBfield}
    \mathbf{\hat{B}}(\mathbf{r},t) &= \frac{i}{c}\sum_{\mathbf{k}, \mu} \left(\frac{\mathbf{k}}{|\mathbf{k}|} \times \mathbf{s}_{\mathbf{k}, \mu} \right) \sqrt{\frac{\hbar \wk}{2 \epsilon_{0} V}} \left(\hat{a}_{\mathbf{k}}(t) e^{i \mathbf{k} \cdot \mathbf{r}} - \hat{a}^{\dagger}_{\mathbf{k}}(t) e^{-i \mathbf{k} \cdot \mathbf{r}}  \right),
\end{align}
\end{subequations}
and,
\begin{align}
    \hat{\mathcal{O}} &= \sqrt{\frac{\hbar}{2 m \omega_{c}}} \left( \hat{b}_{\mathcal{O}} + \hat{b}_{\mathcal{O}}^{\dagger} \right), \\
    \hat{p}_{\mathcal{O}} &= \frac{\sqrt{\hbar m \omega_{c}}}{\sqrt{2} \, i}\left( \hat{b}_{\mathcal{O}} - \hat{b}_{\mathcal{O}}^{\dagger} \right). \quad (\mathcal{O}=x,y)
\end{align}
Here $\wk$ is the frequency of each mode of the EM fields, $V$ is the volume over which the fields have been normalised, $\mathbf{s}_{\mathbf{k}, \mu}$ is the polarisation vector of each mode and $\hat{a}(t) = \hat{a} e^{- i \wk t}$.
The total Hamiltonian $\hat{H}_{\text{tot}}$ then becomes
\begin{align}
    \hat{H}_{\text{tot}} &= \hbar \omega_{c} \left( \hat{b}^{\dagger}_{x} \hat{b}_{x} + \frac{1}{2} \right) + \hbar \omega_{c} \left( \hat{b}^{\dagger}_{y} \hat{b}_{y} + \frac{1}{2} \right) + \sum_{\mathbf{k}, \mu} G_{\mathbf{k}} \; (\mathbf{s}_{\mathbf{k}, \mu} \cdot \mathbf{e}_{x} ) \left( \hat{b}_{x} - \hat{b}^{\dagger}_{x} \right) \left(\hat{a}_{\mathbf{k}}(t) e^{i \mathbf{k} \cdot \mathbf{r}} + \hat{a}^{\dagger}_{\mathbf{k}}(t) e^{-i \mathbf{k} \cdot \mathbf{r}}  \right) \nonumber \\
    &+ \sum_{\mathbf{k}, \mu} G_{\mathbf{k}} \; ( \mathbf{s}_{\mathbf{k}, \mu} \cdot \mathbf{e}_{y} ) \left( \hat{b}_{y} - \hat{b}^{\dagger}_{y} \right) \left(\hat{a}_{\mathbf{k}}(t) e^{i \mathbf{k} \cdot \mathbf{r}} + \hat{a}^{\dagger}_{\mathbf{k}}(t) e^{-i \mathbf{k} \cdot \mathbf{r}}  \right) + \sum_{\mathbf{k}, \mu} \hbar \wk \left( \hat{a}^{\dagger}_{\mathbf{k}} \hat{a}_{\mathbf{k}} + \frac{1}{2} \right),
\end{align}
where,
\begin{align}
    G_{\mathbf{k}} =  \frac{q \hbar}{2 i } \, \sqrt{\frac{\omega_{c}}{\epsilon_{0} m \Omega_{\mathbf{k}} V}}.
\end{align}
We make the long-wavelength approximation
\begin{enumerate}[label=(\roman*)]
\setcounter{enumi}{1}
    \item \label{Ass:appLongWave} \textit{Long-wavelength limit}, $e^{\pm i \kb \cdot \mathbf{r}}\approx 1$.
\end{enumerate}
The total Hamiltonian can now be written
\begin{align}
    \hat{H}_{\text{tot}} = \hat{H}_{x} + \hat{H}_{y} + \hat{H}_{\textsc{f}} + \hat{H}_{\textsc{i}2},
\end{align}
where
\begin{align}
    \hat{H}_{x} &= \hbar \omega_{c} \left( \hat{b}^{\dagger}_{x} \hat{b}_{x} + \frac{1}{2} \right), \label{eq:apphx} \\
    \hat{H}_{y} &= \hbar \omega_{c} \left( \hat{b}^{\dagger}_{y} \hat{b}_{y} + \frac{1}{2} \right), \label{eq:apphy} \\
    \label{eq:apphf}
    \hat{H}_{\textsc{f}} &= \sum_{\mathbf{k}, \mu} \hbar \wk \left( \hat{a}^{\dagger}_{\mathbf{k}} \hat{a}_{\mathbf{k}} + \frac{1}{2} \right),\\
    \nonumber \hat{H}_{\textsc{i}2}&=\sum_{\mathbf{k}, \mu} G_{\mathbf{k}} \; \Bigg\{ ( \mathbf{s}_{\mathbf{k}, \mu} \cdot \mathbf{e}_{x} ) \left( \hat{b}_{x} - \hat{b}^{\dagger}_{x} \right) \left(\hat{a}_{\mathbf{k}}(t) + \hat{a}^{\dagger}_{\mathbf{k}}(t)  \right) \\
    \label{eq:apphi}
    & \hspace*{4 em} + \sum_{\mathbf{k}, \mu} G_{\mathbf{k}} \; ( \mathbf{s}_{\mathbf{k}, \mu} \cdot \mathbf{e}_{y} ) \left( \hat{b}_{y} - \hat{b}^{\dagger}_{y} \right) \left(\hat{a}_{\mathbf{k}}(t) + \hat{a}^{\dagger}_{\mathbf{k}}(t) \right) \Bigg\}.
\end{align}
We study the evolution of $\hat{H}_{\text{tot}}$ through an open quantum system approach outlined in greater detail in Appendix~\ref{app:MEderiv}. The evolution equation for the quatum state $\hat{\sigma}_{xy}$ of the particle in two-dimensional oscillator interacting with the EM vacuum bath is (Eq.~\eref{eq:evoeqa1})
\begin{align}
    \label{eq:appevo}
    \frac{\Delta \widetilde{\sigma}_{xy}(t)}{\Delta t} = -\frac{1}{\hbar^{2}} \frac{1}{\Delta t} \int^{\infty}_{0} d(t_{1} - t_{2}) \int_{t}^{t + \Delta t} d t_{1} \text{Tr}_{\textsc{f}} [ \widetilde{H}_{\textsc{i}2} ( t_{1} ), [ \widetilde{H}_{\textsc{i}2} ( t_{2} ) , \widetilde{\sigma}_{xy} (t) \otimes \widetilde{\sigma}_{\textsc{f}} ]].
\end{align}
Here, we can split the interaction as
\begin{align}
    \widetilde{H}_{\textsc{i}2} = \widetilde{H}_{\textsc{i}x} + \widetilde{H}_{\textsc{i}y}
\end{align}
with
\begin{align}
    \widetilde{H}_{\textsc{i}x} = \sum_{\mathbf{k}, \mu} G_{\mathbf{k}} \; ( \mathbf{s}_{\mathbf{k}, \mu} \cdot \mathbf{e}_{x} ) \left( \widetilde{b}_{x} e^{-i \omega_{c} t} - \widetilde{b}^{\dagger}_{x} e^{i \omega_{c} t} \right) \left(\widetilde{a}_{\mathbf{k}} e^{-i \wk t} + \widetilde{a}^{\dagger}_{\mathbf{k}} e^{i \wk t}  \right), \\
    \widetilde{H}_{\textsc{i}y} = \sum_{\mathbf{k}, \mu} G_{\mathbf{k}} \; ( \mathbf{s}_{\mathbf{k}, \mu} \cdot \mathbf{e}_{y} ) \left( \widetilde{b}_{y} e^{-i \omega_{c} t} - \widetilde{b}^{\dagger}_{y} e^{i \omega_{c} t} \right) \left(\widetilde{a}_{\mathbf{k}} e^{-i \wk t} + \widetilde{a}^{\dagger}_{\mathbf{k}} e^{i \wk t} \right).
\end{align}
These are the interaction terms for $x$ and $y$ respectively with the tilde denoting operators that are now in the Heisenberg picture. Eq.~\eref{eq:appevo} contains a double commutator which results in 4 terms: 
\begin{align}
    \nonumber &\mathcal{C}_{xx} = \text{Tr}_{\textsc{f}}[\widetilde{H}_{\textsc{i}x}(t_{1}),[\widetilde{H}_{\textsc{i}x}(t_{2}),\widetilde{\sigma}_{xy} (t) \otimes \widetilde{\sigma}_{\textsc{f}}]], 
    &\mathcal{C}_{yy}= \text{Tr}_{\textsc{f}}[\widetilde{H}_{\textsc{i}y}(t_{1}),[\widetilde{H}_{\textsc{i}y}(t_{2}),\widetilde{\sigma}_{xy} (t) \otimes \widetilde{\sigma}_{\textsc{f}}]],\\
    &\mathcal{C}_{xy}=\text{Tr}_{\textsc{f}}[\widetilde{H}_{\textsc{i}x}(t_{1}),[\widetilde{H}_{\textsc{i}y}(t_{2}),\widetilde{\sigma}_{xy} (t) \otimes \widetilde{\sigma}_{\textsc{f}}]],
    &\mathcal{C}_{yx}=\text{Tr}_{\textsc{f}}[\widetilde{H}_{\textsc{i}y}(t_{1}),[\widetilde{H}_{\textsc{i}x}(t_{2}),\widetilde{\sigma}_{xy} (t) \otimes \widetilde{\sigma}_{\textsc{f}}]]. \label{eq:doubleCdefn}
\end{align}
Evidently $\mathcal{C}_{xy}$ and $\mathcal{C}_{yx}$ involve cross-products of $\widetilde{H}_{\textsc{i}x}$ and $\widetilde{H}_{\textsc{i}y}$. This is in contrast to $\mathcal{C}_{xx}$ (respectively, $\mathcal{C}_{yy}$) that involves only the product of $\widetilde{H}_{\textsc{i}x}$ (respectively, $\widetilde{H}_{\textsc{i}y}$) with itself. Noting that the summation over $\mathbf{k}$ is transformed to an integral through Eq.~\eref{eq:ksum2},
\begin{equation}
\sum_{\mathbf{k}} f(\mathbf{k}) \rightarrow \frac{V}{8 \pi^{3}}\int_{\mathbf{k}} f(\mathbf{k}) d^{3} \mathbf{k}.
\end{equation}
the commutator
\begin{align}
    \label{eq:Hxy}
    \mathcal{C}_{yx} \propto \int_{0}^{\wmax} \text{d}k \; \int_{0}^{\pi} \text{d} \theta \int_{0}^{2 \pi} \text{d} \phi \; 2|G_{k}|^{2} k^{2} \sin{(\theta)} ( \mathbf{s}_{\mathbf{k}} \cdot \mathbf{e}_{x} ) ( \mathbf{s}_{\mathbf{k}} \cdot \mathbf{e}_{y} ).
\end{align}
Here a factor of 2 accounts for the summation over two independent polarisations $\mu = +, \times$.
Further, we note that the double summation over $\mathbf{k}$ is reduced to a single summation due to the partial trace over the field modes (see Eq.~\eref{eq:doubleCdefn}). Considering $( \mathbf{s}_{\mathbf{k}, \mu} \cdot \mathbf{e}_{x} ) = \sin{(\theta_{\mathbf{k}, \mu})} \cos{(\phi_{\mathbf{k}, \mu})}$ and $( \mathbf{s}_{\mathbf{k}, \mu} \cdot \mathbf{e}_{y} ) = \sin{(\theta_{\mathbf{k}, \mu})} \sin{(\phi_{\mathbf{k}, \mu})}$, we find
\begin{align}
    \mathcal{C}_{yx} \propto \int_{0}^{\wmax} \text{d}k \; \int_{0}^{\pi} \text{d} \theta \int_{0}^{2 \pi} \text{d} \phi \; 2 |G_{k}|^{2} k^{2} \sin^{3}{(\theta)} \cos{(\phi)} \sin{(\phi)} = 0.
\end{align}
Similarly, $\mathcal{C}_{xy}=0$.

The commutator
\begin{align*}
	\mathcal{C}_{xx}= -\sum_{\mathbf{k}, \mu} \sum_{\mathbf{k}', \mu'}  G_{\mathbf{k}} \; G_{\mathbf{k}'}^{\ast} \; ( \mathbf{s}_{\mathbf{k}, \mu} \cdot \mathbf{e}_{x} ) \; ( \mathbf{s}_{\mathbf{k}', \mu'} \cdot \mathbf{e}_{x} ) \left\{ \text{Tr}_{\textsc{f}}\left(  \bigg[ \left( \widetilde{b}_{x} e^{-i \omega_{c} t_{1}} - \widetilde{b}^{\dagger}_{x} e^{i \omega_{c} t_{1}} \right) \widetilde{a}_{\mathbf{k}} e^{-i \wk t_{1}} , \right. \right.\\
	\left.\left. \bigg[\left( \widetilde{b}_{x} e^{-i \omega_{c} t_{2}} - \widetilde{b}^{\dagger}_{x} e^{i \omega_{c} t_{2}} \right) \widetilde{a}^{\dagger}_{\mathbf{k}'} e^{i \Omega_{\kb'} t_{2}} ,\widetilde{\sigma}_{xy} (t) \otimes \widetilde{\sigma}_{\textsc{f}} \bigg] \bigg] + h. c. \right)\right\}.
\end{align*}
On simplification, this yields
\begin{align}
\mathcal{C}_{xx}=\bigg(  &(\bopt_{x} \bdt_{x} \sigmat_{xy} - \bdt_{x} \sigmat_{xy} \bopt_{x} ) \phi^{*}(t_{1} - t_{2})  -    ( \bopt_{x} \bopt_{x} \sigmat_{xy} - \bopt_{x} \sigmat_{xy} \bopt_{x}) \phi^{*}(t_{1}+t_{2})   \nonumber \\
- &  ( \bdt_{x} \bdt_{x} \sigmat_{xy} - \bdt_{x} \sigmat_{xy} \bdt_{x} ) \phi(t_{1}+t_{2})  +   ( \bdt_{x} \bopt_{x} \sigmat_{xy} - \bopt_{x} \sigmat_{xy} \bdt_{x}  ) \phi(t_{1}-t_{2})  \bigg) \chi_{\textsc{f}1} \nonumber \\
+ & \bigg(  -( \bopt_{x} \sigmat_{xy} \bdt_{x} - \sigmat_{xy} \bdt_{x} \bopt_{x} ) \phi^{*}(t_{1}-t_{2})   +  ( \bopt_{x} \sigmat_{xy} \bopt_{x} - \sigmat_{xy} \bopt_{x} \bopt_{x} ) \phi^{*}(t_{1}+t_{2})   \nonumber \\
+ &  ( \bdt_{x} \sigmat_{xy} \bdt_{x} - \sigmat_{xy} \bdt_{x} \bdt_{x} ) \phi(t_{1}+t_{2}) -  ( \bdt_{x} \sigmat_{xy} \bopt_{x} - \sigmat_{xy} \bopt_{x} \bdt_{x} ) \phi(t_{1}-t_{2})   \bigg)  \chi_{\textsc{f}2}  +  h.c,
\end{align}
where
\begin{align}
\chi_{\textsc{f}1}&=\sum_{\mathbf{k}, \mu}  |G_{\mathbf{k}} ( \mathbf{s}_{\mathbf{k}, \mu} \cdot \mathbf{e}_{x} )|^{2} e^{-i \wk (t_{1}-t_{2})} \bigg( \aver{n_{\kb}} + \aver{[\widetilde{a}_{\mathbf{k}},\widetilde{a}_{\mathbf{k}}^{\dagger}]} \bigg),\\
\chi_{\textsc{f}2}&=\sum_{\mathbf{k}, \mu}  |G_{\mathbf{k}} ( \mathbf{s}_{\mathbf{k}, \mu} \cdot \mathbf{e}_{x} )|^{2} e^{i \wk (t_{1}-t_{2})}  \aver{n_{\kb}},\\
\phi(t) &= e^{i \omega_{c} t}.
\end{align}
The commutator $\mathcal{C}_{yy}$ is obtained similarly.

With Eq.\eref{eq:appevo} becoming
\begin{align*}
\frac{\Delta \widetilde{\sigma}_{xy}(t)}{\Delta t} = -\frac{1}{\hbar^{2}} \frac{1}{\Delta t} \int^{\infty}_{0} d(t_{1} - t_{2}) \int_{t}^{t + \Delta t} d t_{1} \left( \mathcal{C}_{xx} +\mathcal{C}_{yy} \right),
\end{align*}
and on evaluating the integrals using Eq. \eref{eq:ints},  the master equation simplifies to
\begin{align}
\frac{\Delta \widetilde{\sigma}_{xy}}{\Delta t} =  & \left( \bdt_{x} \sigmat_{xy} \bopt_{x} - \bopt_{x} \bdt_{x} \sigmat_{xy} - \zeta(t)  ( \bdt_{x} \sigmat_{xy} \bdt_{x} - \bdtp{2}_{x} \sigmat_{xy} ) \right) \left( \frac{1}{2}\Gamma'_{+} + \frac{1}{2}\Gamma_{+}  - i\Delta_{+} - i\Delta'_{+} \right)   \nonumber \\
+ &  \left( \bopt_{x} \sigmat_{xy} \bdt_{x} - \bdt_{x} \bopt_{x} \sigmat_{xy} - \zeta^{*}(t)  ( \bopt_{x} \sigmat_{xy} \bopt_{x} - \bopt^{2}_{x} \sigmat_{xy} ) \right) \left( \frac{1}{2}\Gamma'_{-} + \frac{1}{2}\Gamma_{-}  + i\Delta_{-} + i\Delta'_{-} \right) \nonumber \\
+ &  \left(\bopt_{x} \sigmat_{xy} \bdt_{x} - \sigmat_{xy} \bdt_{x} \bopt_{x} - \zeta(t)   ( \bdt_{x} \sigmat_{xy} \bdt_{x} - \sigmat_{xy} \bdtp{2}_{x}  ) \right) \left( \frac{1}{2}\Gamma'_{+} - i\Delta'_{+} \right) \nonumber \\
+ &  \left( \bdt_{x} \sigmat_{xy} \bopt_{x} - \sigmat_{xy} \bopt_{x} \bdt_{x}  - \zeta^{*}(t)  ( \bopt_{x} \sigmat_{xy} \bopt_{x} - \sigmat_{xy} \bopt^{2}_{x} ) \right) \left( \frac{1}{2}\Gamma'_{-} + i\Delta'_{-} \right) \nonumber\\
+& \left( \bdt_{y} \sigmat_{xy} \bopt_{y} - \bopt_{y} \bdt_{y} \sigmat_{xy} - \zeta(t)  ( \bdt_{y} \sigmat_{xy} \bdt_{y} - \bdtp{2}_{y} \sigmat_{xy} ) \right) \left( \frac{1}{2}\Gamma'_{+} + \frac{1}{2}\Gamma_{+}  - i\Delta_{+} - i\Delta'_{+} \right)   \nonumber \\
+ &  \left( \bopt_{y} \sigmat_{xy} \bdt_{y} - \bdt_{y} \bopt_{y} \sigmat_{xy} - \zeta^{*}(t)  ( \bopt_{y} \sigmat_{xy} \bopt_{y} - \bopt^{2}_{y} \sigmat_{xy} ) \right) \left( \frac{1}{2}\Gamma'_{-} + \frac{1}{2}\Gamma_{-}  + i\Delta_{-} + i\Delta'_{-} \right) \nonumber \\
+ &  \left(\bopt_{y} \sigmat_{xy} \bdt_{y} - \sigmat_{xy} \bdt_{y} \bopt_{y} - \zeta(t)   ( \bdt_{y} \sigmat_{xy} \bdt_{y} - \sigmat_{xy} \bdtp{2}_{y}  ) \right) \left( \frac{1}{2}\Gamma'_{+} - i\Delta'_{+} \right) \nonumber \\
+ &  \left( \bdt_{y} \sigmat_{xy} \bopt_{y} - \sigmat_{xy} \bopt_{y} \bdt_{y}  - \zeta^{*}(t)  ( \bopt_{y} \sigmat_{xy} \bopt_{y} - \sigmat_{xy} \bopt^{2}_{y} ) \right) \left( \frac{1}{2}\Gamma'_{-} + i\Delta'_{-} \right) + h.c.
\end{align}
Here we recall the phase factor $\zeta$ (Eq. \eref{eq:zetaDefn})
\begin{align}
    \zeta(t) = \int_{t}^{t + \Delta t} d t_{1} \; \left[\phi(t_{1})\right]^{2} = \frac{e^{i 2 \omega_{c} t}( e^{2 i \omega_{c} \Delta t} - 1 )}{2 i \omega_{c} \Delta t}.
\end{align}
Further, noting that $( \mathbf{s}_{\mathbf{k}, \mu} \cdot \mathbf{e}_{x} )=\sin \theta_{\kb,\mu} \cos \phi_{\kb,\mu}$, $( \mathbf{s}_{\mathbf{k}, \mu} \cdot \mathbf{e}_{y} )=\sin \theta_{\kb,\mu} \sin \phi_{\kb,\mu}$, and 
\begin{equation}
\int_{0}^{2 \pi} \rmd \phi_{\kb,\mu} \cos^{2} \phi_{\kb,\mu}= \int_{0}^{2 \pi} \rmd \phi_{\kb,\mu} \sin^{2} \phi_{\kb,\mu}= \frac{1}{2}, 
\end{equation}
we have
\begin{subequations}
\begin{align}
\label{eq:c1Cyc}
\Gamma_{\pm} &\equiv \sum_{\kb} \frac{|G_{\kb} ( \mathbf{s}_{\mathbf{k}, \mu} \cdot \mathbf{e}_{x} )|^{2}}{\hbar^{2}} 2 \pi \, \delta \left( \wk \pm \omega_{c} \right) \: \langle [\aop, \ad] \rangle =  \sum_{\kb} \frac{|G_{\kb} ( \mathbf{s}_{\mathbf{k}, \mu} \cdot \mathbf{e}_{y} )|^{2}}{\hbar^{2}} 2 \pi \, \delta \left( \wk \pm \omega_{c} \right) \: \langle [\aop, \ad] \rangle, \\
\label{eq:c2Cyc}
\Gamma_{\pm}^{'} &\equiv \sum_{\kb} \frac{|G_{\kb} ( \mathbf{s}_{\mathbf{k}, \mu} \cdot \mathbf{e}_{x} )|^{2}}{\hbar^{2}} 2 \pi \, \delta \left( \wk \pm \omega_{c} \right) \:  \langle n_{k} \rangle = \sum_{\kb} \frac{|G_{\kb} ( \mathbf{s}_{\mathbf{k}, \mu} \cdot \mathbf{e}_{y} )|^{2}}{\hbar^{2}} 2 \pi \, \delta \left( \wk \pm \omega_{c} \right) \:  \langle n_{k} \rangle,  \\
\label{eq:c3Cyc}
\Delta_{\pm} &\equiv \sum_{\kb} \frac{|G_{\kb} ( \mathbf{s}_{\mathbf{k}, \mu} \cdot \mathbf{e}_{x} )|^{2}}{\hbar^{2}} \frac{1}{\omega_{c} \pm \wk} \langle [\aop, \ad] \rangle = \sum_{\kb} \frac{|G_{\kb} ( \mathbf{s}_{\mathbf{k}, \mu} \cdot \mathbf{e}_{y} )|^{2}}{\hbar^{2}} \frac{1}{\omega_{c} \pm \wk} \langle [\aop, \ad] \rangle,  \\
\label{eq:c4Cyc}
\Delta_{\pm}^{'} &\equiv \sum_{\kb} \frac{|G_{\kb} ( \mathbf{s}_{\mathbf{k}, \mu} \cdot \mathbf{e}_{x} )|^{2}}{\hbar^{2}} \frac{1}{\omega_{c} \pm \wk} \langle n_{k} \rangle = \sum_{\kb} \frac{|G_{\kb} ( \mathbf{s}_{\mathbf{k}, \mu} \cdot \mathbf{e}_{y} )|^{2}}{\hbar^{2}} \frac{1}{\omega_{c} \pm \wk} \langle n_{k} \rangle.
\end{align} 
\label{eqn:csetCyc}
\end{subequations}
It is important to emphasize that we have intentionally used the same notation for Eq. \eref{eqn:csetCyc} in the context of the cyclotrons and Eq. \eref{eqn:constset} in the context of the one-dimensional oscillator. This is becasue on evaluating Eq. \eref{eqn:csetCyc} by using Eq. \eref{eq:ksum2}, it would be identical to the expression obtained from Eq. \eref{eqn:constset}. 

As in Appendix \ref{app:MEderiv}, we obtain the master equation by taking the limit $\Delta t \rightarrow 0$, setting $\langle n_{\kb} \rangle = 0$, moving to the Schr{\"o}dinger picture, and evaluating Eqs. \eref{eq:c1Cyc}-\eref{eq:c4Cyc} using Eq. \eref{eq:ksum2}. We also use the notation $\Gamma\equiv\Gamma_{-}$. The master equation is 
\begin{align}
    \frac{\text{d} \hat{\sigma}_{xy}}{\text{d}t} = - &  i \left( \omega_{c} + \Delta_{-} - \Delta_{+} \right) \left[ \bxd \bxop , \hat{\sigma}_{xy} \right]-  i \left( \omega_{c} + \Delta_{-} - \Delta_{+} \right) \left[ \byd \byop , \hat{\sigma}_{xy} \right] \nonumber \\
    + & \Gamma \left( \bxop \hat{\sigma}_{xy} \bxd - \frac{1}{2}\bxd \bxop \hat{\sigma}_{xy} - \frac{1}{2}\hat{\sigma}_{xy} \bxd \bxop \right) \nonumber \\ 
    + & \Gamma \left( \byop \hat{\sigma}_{xy} \byd - \frac{1}{2}\byd \byop \hat{\sigma}_{xy} - \frac{1}{2}\hat{\sigma}_{xy} \byd \byop \right) \nonumber \\ 
    - & i \Delta_{+} \left( \bxop \hat{\sigma}_{xy} \bxop - \hat{\sigma}_{xy} \bxop^{2} \right) \nonumber - \left( \frac{1}{2}\Gamma + i \Delta_{-} \right) \left( \bxop \hat{\sigma}_{xy} \bxop - \bxop^{2} \hat{\sigma}_{xy} \right)  \nonumber \\ 
    - & i \Delta_{+} \left( \byop \hat{\sigma}_{xy} \byop - \hat{\sigma}_{xy} \byop^{2} \right) \nonumber - \left( \frac{1}{2}\Gamma + i \Delta_{-} \right) \left( \byop \hat{\sigma}_{xy} \byop - \byop^{2} \hat{\sigma}_{xy} \right)  \nonumber \\ 
    + &  i \Delta_{+} \left( \bxd \hat{\sigma}_{xy} \bxd -  \bdp{2}_{x} \hat{\sigma}_{xy} \right) - \left( \frac{1}{2}\Gamma -  i \Delta_{-} \right) \left( \bxd \hat{\sigma}_{xy} \bxd - \hat{\sigma}_{xy} \bdp{2}_{x} \right) \nonumber \\
    + &  i \Delta_{+} \left( \byd \hat{\sigma}_{xy} \byd -  \bdp{2}_{y} \hat{\sigma}_{xy} \right) - \left( \frac{1}{2}\Gamma -  i \Delta_{-} \right) \left( \byd \hat{\sigma}_{xy} \byd - \hat{\sigma}_{xy} \bdp{2}_{y} \right).
    \label{eqn:CycME}
\end{align}
where $\Delta_{\pm}$ and $\Gamma$ are as defined in Eqs. \eref{eq:deltapm2} and \eref{eq:GammaPM} respectively. Defining $\sigma_{x}$ as $\text{Tr}_{y} [ \sigma_{xy} ]$ and $\sigma_{y}$ as $\text{Tr}_{x} [ \sigma_{xy} ]$, the master equation can be split into two independent master equations for $x$ and $y$.
\begin{align}
    \frac{\text{d} \hat{\sigma}_{xy}}{\text{d}t} = \frac{\text{d} \hat{\sigma}_{x}}{\text{d}t} + \frac{\text{d} \hat{\sigma}_{y}}{\text{d}t}.
\end{align}
where
\begin{align}
    \label{eq:mastereqx}
    \frac{\text{d} \hat{\sigma}_{x}}{\text{d}t} = \text{Tr}_{y} \left[ \frac{\text{d} \hat{\sigma}_{xy}}{\text{d}t} \right] = & - i \left( \omega_{c} + \Delta_{-} - \Delta_{+} \right) \left[ \bd_{x} \bop_{x} , \hat{\sigma}_{x} \right] \nonumber \\
    + & \Gamma \left( \bop_{x} \hat{\sigma}_{x} \bd_{x} - \frac{1}{2} \bd_{x} \bop_{x} \hat{\sigma}_{x} - \frac{1}{2} \hat{\sigma}_{x} \bd_{x} \bop_{x} \right) \nonumber \\ 
    + & i \Delta_{+} \left( \bd_{x} \hat{\sigma}_{x} \bd_{x} - \bdp{2}_{x} \hat{\sigma}_{x} - \bop_{x} \hat{\sigma}_{x} \bop_{x} + \hat{\sigma}_{x} \bop_{x}^{2} \right) \nonumber \\
    + & i \Delta_{-} \left( \bd_{x} \hat{\sigma}_{x} \bd_{x} - \hat{\sigma}_{x} \bdp{2}_{x} - \bop_{x} \hat{\sigma}_{x} \bop_{x} + \bop^{2}_{x} \hat{\sigma}_{x} \right) \nonumber \\
    + & \frac{1}{2} \Gamma \left( \hat{\sigma}_{x} \bdp{2}_{x} - \bd_{x} \hat{\sigma}_{x} \bd_{x} - \bop_{x} \hat{\sigma}_{x} \bop_{x} + \bop^{2}_{x} \hat{\sigma}_{x} \right). 
\end{align}
A similar equation to Eq.~\eref{eq:mastereqx} can be found for the $y$ motion by swapping the the subscripts $x \leftrightarrow y$.

Comparison of Eq.~\eref{eq:mastereqx} and Eq.~\eref{eq:mastereq2} show that the master equation for a one-dimensional oscillator is identical to the master equation for the $x$-axis (or $y$) motion of a two-dimensional oscillator. We therefore apply the results obtained through analysis of the one-dimensional oscillator to the two-dimensional motion of the SEC.

\end{document}